\documentclass[sigconf]{acmart}

\definecolor{darkgreen}{rgb}{0.0, 0.6, 0.2} 
\definecolor{lightgreen}{rgb}{0.3, 0.5, 0.4} 

\AtBeginDocument{%
  }

\usepackage{float}
\usepackage{fontawesome}
\usepackage{multirow}
\definecolor{Gallery}{rgb}{0.937,0.937,0.937}
\usepackage{colortbl}
\usepackage{soul}
\usepackage{enumitem}
\usepackage{fp}
\usepackage{subcaption}
\usepackage[dvipsnames]{xcolor}
\usepackage{setspace}
\usepackage{graphicx}
\usepackage{tabularx}
\usepackage{setspace}
\usepackage{booktabs}
\usepackage{makecell}
\usepackage{xspace} 
\usepackage{multirow}
\usepackage{cleveref}
\usepackage{orcidlink}

\usepackage[final, commandnameprefix=ifneeded, commentmarkup=margin, defaultcolor=blue]{changes}
\setcommentmarkup{} 
\definechangesauthor[name=Sadia]{SA}
\definechangesauthor[name=Rudrajit]{RC}
\definechangesauthor[name=Fatima]{FAM}
\definechangesauthor[name=Amreeta]{AC}
\definechangesauthor[name=Margaret]{MMB}
\definechangesauthor[name=Anita]{AS}

\newif\ifdraft
\draftfalse 

\newcolumntype{C}{>{\centering\arraybackslash}X}

\newcommand{\boldification}[1]{\ifdraft%
    **\textbf{#1}** 
     \fi
}

\newcommand{\draftStatus}[1]{\ifdraft%
    **\textbf{Status: #1}**\else\relax\fi
}

\newcommand{\fixme}[2]{\ifdraft%
    \par\noindent%
    {
    \color{red}\textbf{**FIXME** #1: #2}
    }%
    \fi
}

\definecolor{outputGreen}{HTML}{90bd17}
\definecolor{userBlue}{HTML}{82aff5}

\usepackage[breakable]{tcolorbox}

\newcommand{\mybox}[1]{\par\noindent
  \begin{tcolorbox}[breakable, colback=gray!25, colframe=gray!25,
                    boxrule=0pt, arc=0pt, boxsep=0pt,
                    left=\fboxsep, right=\fboxsep,
                    top=\fboxsep, bottom=\fboxsep]
  \indent #1
  \end{tcolorbox}}

\newcommand{\borderedbox}[1]{%
\begin{singlespace}
\small\texttt{%
\fbox{%
  \parbox{\dimexpr\linewidth-10pt-2\fboxsep-2\fboxrule\relax}{%
    #1
  }%
}%
}
\end{singlespace}
}

\acmISBN{000-0-0000-XXXX-X/2027/04}

\begin{document}

\title{Apply-\texorpdfstring{$<$x$>$}{<x>}Mag: \\One Tool to Support Many Inclusive Design Methods}



\author{
Sadia Afroz\orcidlink{0000-0002-2485-548X}$^1$ \quad
Rudrajit Choudhuri\orcidlink{0000-0001-7168-2107}$^1$ \quad
Fatima A. Moussaoui\orcidlink{0009-0005-0542-0324}$^1$ \quad
Amreeta Chatterjee\orcidlink{0000-0003-1902-8313}$^2$ \\
Margaret Burnett\orcidlink{0000-0001-6536-7629}$^1$ \quad
Anita Sarma\orcidlink{0000-0002-1859-1692}$^1$
}
\affiliation{
$^1$Oregon State University, OR, USA.
Email: \{afrozs, choudhru, moussaof, anita.sarma\}@oregonstate.edu, burnett@eecs.oregonstate.edu \\
$^2$University at Albany, NY, USA. 
Email: achatterjee3@albany.edu
}
\renewcommand{\shortauthors}{Afroz et al.}







\begin{abstract}
\draftStatus{MMB 9/9/26: D3}
\boldification{what's the problem}
Doing inclusive design in HCI practice can be labor-intensive, a costly barrier that some companies and HCI practitioners may be unwilling or unable to overcome. 
\boldification{why is the problem a problem}
Yet, \textit{not} doing inclusive design is costly too, in the form of UX barriers that disproportionately disadvantage under-served user populations. 
\boldification{what did we do about it}
To address this problem, we introduce Apply-<x>Mag, an LLM-powered tool to support HCI practitioners' work to design their products inclusively to wide ranges of users. 
\boldification{what does this do for the world (some fantastic results)}
Apply-<x>Mag is general, supporting \textit{any} inclusive design method that can be expressed as <x>Mags (i.e., using attribute ranges and heuristics).  
It is also effective: Empirical results with researcher and practitioner teams using various combinations of two <x>Mags on 7 products showed Apply-<x>Mag precision averaging 90--99\% and recall averaging 82--89\%. 
Further, its environmental costs were reasonable, costing 
about the same resources as 2--4 ordinary Google searches. 

\end{abstract}





\settopmatter{
    printacmref=false,
    printccs=false,
    printfolios=true} 
\renewcommand\footnotetextcopyrightpermission[1]{} 

\makeatletter
\let\ACM@savedClassError\ClassError
\def\ClassError#1#2#3{\ClassWarning{#1}{#2}}
\makeatother

\maketitle

\makeatletter
\let\ClassError\ACM@savedClassError
\makeatother

\ifdraft
\input{sections/00-globalTODOs}
     \fi
\section{Introduction\draftStatus{MMB 9/3: d2.5}}
\label{sec:intro}

\boldification{-----1. what's the problem: inclusivity matters for, among other reasons, market share. But doing inclusivity work can be time-consuming and expensive.}

Many people feel overlooked by current technologies' designs and workflows, and with good reason---technology practitioners often do inadvertently overlook wide swaths of potential user populations~\cite{harrington2022blackOlderAdults, zallio2022designingMetaverse, liao2022AItrust, mubarak2022elderly, radanliev2024accessibility}.
Inclusive design methods exist to help technology practitioners avert this problem, but using these methods can be more labor-intensive than some companies can afford.
Further, some companies may lack \added[id=RC]{relevant} expertise to use these methods well.

\boldification{1 (cont). what's the problem: some methods have tools to help with this to some extent (which we cover in related work), but we propose a more powerful approach that goes beyond a 1-1 relationships between methods \& tools to a many-to-1 relationship: 1 tool for many methods.}
A partial solution to the labor-intensiveness of many inclusive design methods is tools that partially/fully automate or scaffold such methods.
Some inclusive design methods have such tools, as we discuss further in Section~\ref{sec:background+Related}, but hand-building a new tool for every method is expensive to create and expensive to maintain as the method evolves.
To address this problem, we propose to move beyond one-tool-one-method relationships, to a one-to-many relationship---\textit{one} tool to automate \textit{many} inclusive design methods.

\boldification{1 (cont). what's the problem: what's inclusive design, what are inclusivity bugs}

In this paper, we use the term ``inclusive design method'' broadly, to refer to any approach to technology design that aims to improve the product's user experience quality for population groups it did not serve well before.%
\footnote{Definitions of inclusive design vary in the literature (e.g.,~\cite{oneill2021accessibility, patrick2021design-for-all, persson2015universalDesign}). Our use of the term is deliberately broad, and does not differentiate between inclusive design and near-synonyms such as design-for-all and universal design. To paraphrase Kat Holmes's simple definition~\cite{holmes2018mismatch}, inclusive design is the opposite of exclusionary design.
} 
In keeping with this definition, the term ``inclusivity bug'' refers to usability bugs that \textit{disproportionately} disadvantage some particular group of users.
For example, inclusivity bugs in a widely used class registration system arose disproportionately often for lower-socioeconomic students~\cite{busteed2026}.

\boldification{Tool handles any xMag.  To do this we harness the power of LLMs.}

To automatically find and recommend fixes to such inclusivity bugs, we present Apply-<x>Mag, a new inclusive design tool that supports HCI practitioners in improving their products' inclusiveness for a wide range of users.
\added{
Apply-<x>Mag is a generalized tool that can automate any inclusive design method in the InclusiveMag family~\cite{mendez2019inclusivemag} (any ``<x>Mag'').
As we discuss in Section~\ref{subsec:background}, the family currently has eight <x>Mags, the best known of which is GenderMag~\cite{burnett2016gendermag}.
The tool targets automating these eight, as well as any other inclusive design method expressible using attribute ranges and heuristics, by harnessing the power of LLMs.
}

\boldification{-----3. what's hard about it/what are the pieces of it/what are we aiming to accomplish}
A tool with the generality we envision must fulfill five requirements. 
\textit{Generality}: It must be able to apply any <x>Mag (past, present, or future) without hand-coding. Section~\ref{sec:tool} explains how we fulfilled this requirement.
\textit{Fidelity to the method}: The tool must stay true to the <x>Mag the practitioner selects, avoiding ``goal drift'' into stereotypes and pseudo-guidelines external to the method. Section~\ref{sec:eval-design} empirically evaluates this goal by comparing the tool's results against humans applying the same inclusive design method manually.
\textit{Viability for HCI practice}: It must not  require an HCI practitioner to enter information they would not necessarily have, such as a list of known inclusivity bugs to ``train'' the tool on a particular <x>Mag. Section~\ref{subsec:Use-case2} shows how the tool fulfills this requirement.
\textit{Flexibility in HCI practice}: It must enable HCI practitioners to pick and choose (portions of) <x>Mags to run together, such as choosing only certain elements of one <x>Mag or running multiple <x>Mags together for intersectional insights. Section~\ref{subsec:Use-case4} shows how the tool fulfills this requirement.
\textit{Environmentally responsible}: The tool harnesses an LLM, and LLM usage can be costly to the environment. In Sections~\ref{subsec:results-environment} and~\ref{sec:discussion} we compare the environmental cost of using Apply-<x>Mag against chatting with an LLM  directly, and then go on to consider the environmental and other costs of using our tool versus using an interactive LLM versus doing the work manually. 


\boldification{-----5. Our contributions are:}

Thus, this paper contributes the following:

\begin{enumerate}
     \item Apply-<x>Mag, a tool to support HCI practitioners' inclusive design work by automating:
        \begin{itemize}
        \item any <x>-Mag (current or future);
        \item any subset or variant (current or future);
        \item any intersectional combination of the above.
        \end{itemize}
    \item Empirical results of Apply-<x>Mag's effectiveness evaluating 7 technology products (Section~\ref{sec:eval-design}).
    \item Empirical comparison of using Apply-<x>Mag vs. an LLM chat vs. manual <x>Mags (Section~\ref{subsec:results-environment} and Section~\ref{sec:discussion}). 
\end{enumerate}



\paragraph{\textbf{Positionality}.\draftStatus{d2: MMB 8/24/26}}
We are of multiple races (Asian, White, MENA), with national/ethnic backgrounds from Asian, Middle Eastern/North African, and North American nations. Several of us also hold the intersectional identity of women of color. A number of us have personally experienced underrepresentation in computing. At the same time, we recognize the privileges we hold as individuals with access to higher education and to the resources required to conduct this research. We approach this work as a contribution toward easing designers' and software practitioners' path to creating inclusive technology, not a comprehensive remedy, and present our findings with that scope in mind.

\section{Background and Related Work}
\label{sec:background+Related}

\subsection{Background: How the InclusiveMag family of methods work\draftStatus{D2.8 AC 9/9/26}}
\label{subsec:background}


\boldification{Our tool automates present and future inclusive design methods created by the InclusiveMag meta-method~\cite{mendez2019inclusivemag}. InclusiveMag is...}
The Apply-<x>Mag tool automates practitioners' application of methods previously created by inclusive design researchers using InclusiveMag~\cite{mendez2019inclusivemag}.
InclusiveMag is a meta-method---a method for creating new methods.
Specifically, it supports inclusive design researchers in creating new, analytical inclusive design methods to support a diversity dimension of their choice (e.g., to support gender diversity, age diversity, ...).
The tool does not automate InclusiveMag per se; it automates practitioners' downstream \textit{use} of the ``<x>Mag'' inclusive design method that researchers have used InclusiveMag to create.

\begin{figure} [h]
    \centering
    \includegraphics[width=0.9\linewidth]{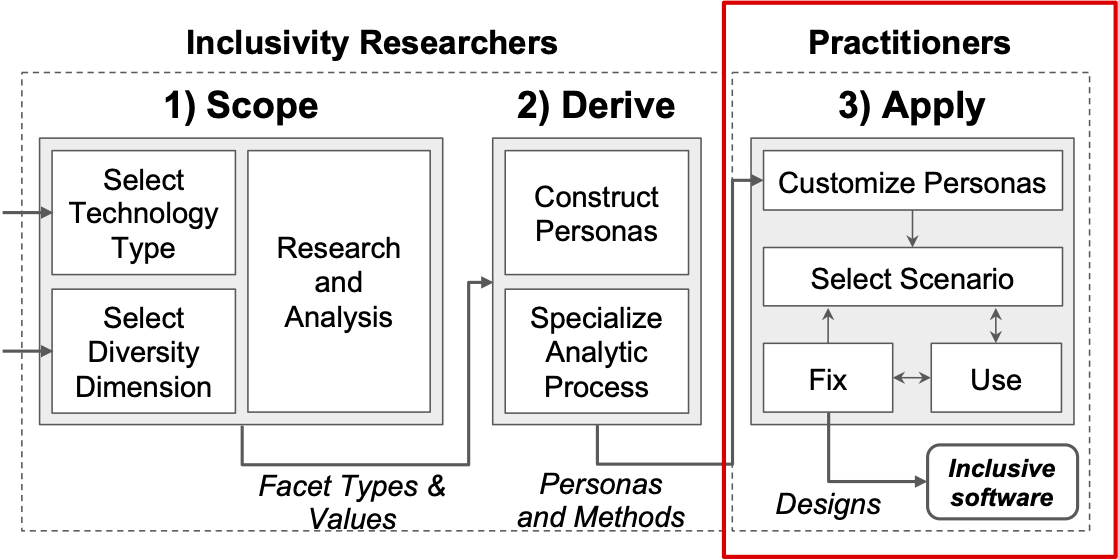}
    \caption{
    The InclusiveMag meta-method~\cite{mendez2019inclusivemag} has three steps, each having multiple components. Inclusive design researchers create a new inclusive design method (a new ``<x>Mag'') for the diversity dimension \textit{x} they choose via Steps-1 (Scope) and -2 (Derive). In Step-3 (Apply, in superimposed \textcolor{red}{red box}), practitioners apply the <x>Mag resulting from Steps 1--2 to their products. Apply-<x>Mag automates Step-3.}
    \label{fig:inclusivemag}
    \Description{A flowchart depicting the InclusiveMag meta-method. There are three boxes, one for each step. The first box, labeled Step-1 (Scope), has three components: Select Technology Type, Select Diversity Dimension, and Research and Analysis. An arrow (labeled Facet Types \& Values) points from the first box to the second, labeled Step-2 (Derive), which has two components: Construct Personas, and Specialize Analytic Process. An arrow (labeled Personas and Methods) points from the second box to the third box, labeled Step-3 (Apply), which has four components: Customize Personas, Select Scenario, Use, and Fix. Customize Personas has a one-directional arrow pointing to Select Scenario. Select Scenario has a two-directional arrow between itself and Use. Use has a two-directional arrow between itself and Fix. Fix has a one-directional arrow that points back to Select Scenario, creating a loop, but has another arrow (labeled Designs) pointing outside the box to the final output: Inclusive software.
}
\end{figure}

\boldification{Figure~\ref{fig:inclusivemag} shows how InclusiveMag works. In step-1, researchers set the scope and figure out facet types \& values.}
Figure~\ref{fig:inclusivemag} shows how the InclusiveMag meta-method works to enable researchers to create a new <x>Mag method for some type of diversity (\textit{x}).
Researchers manually do Steps-1 and -2 to create a new <x>Mag before practitioners can use it in Step-3.
Here we describe the researchers' work to create such <x>Mags to provide a theoretical understanding of how these methods work.

In Step-1 (\textit{Scope}), researchers set the <x>Mag method's scope by choosing a technology type and a diversity dimension \textit{x}.
For SocioeconomicMag (abbreviated SESMag), \textit{x} was socioeconomic-status (SES) and the genre was problem-solving software~\cite{burnett2024SESMag, busteed2026}.
%
%
They then identify attributes that research shows differ substantially across the dimension (e.g., with SESMag, differing attitudes toward authority clustering around lower- vs. higher-SES individuals). 
A subset of these attributes become facet types (as per criteria the researchers choose), with attribute endpoints defining each facet type's range of values.
\footnote{InclusiveMag facets all need to be (approximately) ordinal.
} 


%

\boldification{The facet endpoint values matter!}
These facet value endpoints are critical to the practicality of the <x>Mag inclusive design methods.
The InclusiveMag philosophy is that, when an inclusivity bug is unearthed by an <x>Mag, the fix is to design features/workflows that serve both endpoints of the facet types \textit{simultaneously}, because doing so covers not only users with exactly those endpoint facet values, but also users in between those endpoints, and users whose facet values change dynamically.
The downstream impact on practitioners who eventually apply an <x>Mag to their own products is that they need to do <x>Mag sessions only twice on their scenarios---once for the first set of endpoints and again for the second.

\boldification{In step-2 these endpoints turn into personas, and...}

In Step-2 (\textit{Derive}), researchers turn the facet types and values into the new <x>Mag method. Most define at least two personas to bring the facet types and values to life. One persona represents the collection of ``low'' endpoints of each facet type's range, and another the collection of the ``high'' endpoints. SESMag's Dav persona (Figure ~\ref{fig:dav}, shown later) is an example of SESMag facet endpoints common among lower-SES individuals; an ``opposite'' SESMag persona, Fee, is defined by the endpoints common among higher-SES individuals. 
Additional personas with any desired mix of the facet values are possible, but the two endpoint personas suffice: they span the range of facet values without persona explosion.  
The facet values completely define each persona.

The researcher then specializes an analytic process such as a specialized cognitive walkthrough or set of specialized heuristics~\cite{steine2019fixing, burnett2024SESMag}, to produce a concrete inclusive design method (in this example, SESMag).
The facet values and the specialized analytic process thus completely define the <x>Mag.
\fixme{mmb}{NEVER delete above sentence about the facets+process completely defining the Mag.  Also this sentence needs to remain as the LAST SENTENCE in the PARAGRPH.)}

The Apply-<x>Mag tool we present in this paper comes into play in Step-3.
In Step-3 (\textit{Apply}), practitioners apply <x>Mag to their own product using the analytic process the researchers specialized in Step-2. Using an endpoint persona, the practitioner walks through a use-case or scenario one action at a time, and at each action answers the <x>Mag's specialized walkthrough questions using the persona's facet values: whether the persona would do that action, and would recognize from the software's feedback that they had made progress. Wherever the persona would falter, the practitioner has found an inclusivity bug.
Apply-<x>Mag replaces this walkthrough with an automated heuristic evaluation based on the facets.

\boldification{Inclusive design researchers have used InclusiveMag to lots of Mags, used around the world}
Inclusive design researchers have used InclusiveMag to create eight inclusive design methods so far: GenderMag~\cite{burnett2016gendermag}, 
GenderMag-for-AI~\cite{anderson_over-the-hood_2026}
AgeMag~\cite{mcintosh2021agemag}, 
HealthMag~\cite{xiao2026elderlyHealthmag},
Elderly HealthMag~\cite{xiao2026elderlyHealthmag}, 
RemoteCollabEval~\cite{mason2026equity}, 
SocioeconomicMag (abbreviated SESMag)~\cite{burnett2024SESMag, busteed2026, chikezie2025SESmagSurvey}, 
and Fallatah et al.’s Intersectional HCI approach~\cite{fallatah2025intersectionalmag}.  
Perhaps the best known of these is GenderMag, which has been adopted by technologists across the world (e.g., ~\cite{burnett2016field, 
vorvoreanu2019genderMag, 
murphyhill2024icse, 
obrien2025genderMaginEducation, 
culas2025genderMagnewcomers, 
shekhar2018cognitive, 
cunningham2016genderMag, 
barbosa2021diversidade, 
zanardi2025GenderMagkids, 
aveiro2025genderMaglowcode}). 
<x>Mag methods like these become configurations for the Apply-<x>Mag tool, as later sections will show.


\subsection{Related work\draftStatus{D2.75 (MMB) 9/9/26}}



\boldification{Automated support for finding usability problems is well established. Among those, finding accessibility problems is the most popular kind of tools}

Automated interface inspection has been studied since the 1990s, when the earliest work set out to automate conformance to accessibility standards~\cite{balbo1995software, ivory2001state}. That effort has since matured into a wide range of tools: WAVE~\cite{kasday2000tool} checks web pages for accessibility barriers, Vischeck~\cite{dougherty2000vischeck} simulates low vision, AATT~\cite{khan2021aatt} tests for WCAG conformance, and Ally~\cite{pradhan2022development} improves the accessibility of PDF files. Since then, a substantial body of work has automated accessibility evaluation (with and without LLMs)~\cite{ara2025inclusive, duarte2025expanding, eler2018automated, lempola2024comparing, salehnamadi2022groundhog, uwase2026comparative}. This work is mature and widely used, but it tests conformance to an accessibility standard. Inclusivity across a wide range of populations, which is our tool's aim, is a different question.


\boldification{There is a whole host literature about customizing to \emph{just 1} specific user, our goal with Apply-<x>Mag is to support practitioners in making their interfaces work for lots of people}

A substantial line of research on personalization seeks to support human differences by customizing the interface to \emph{just one} user~\cite{kim2022stylette, li2023using, kong2024ability, kong2025supporting, alves2024citizen, cao2025generative, alves2026exploring}. In essence, the philosophy of personalization systems is ``one size fits one''.
Recent instances configure accessible visualizations for the individual~\cite{jones2024customization}
and teach personalized accessibility systems one user at a time \cite{wen2024find}. Another branch of this work monitors a user's behavior and predicts what the interface should surface for them \cite{khamaj2024adapting, zhan2024personalized}. 

In contrast to ``one size fits one'' approaches like personalization, inclusive and universal design aim for ``one size fits all''~\cite{mace1991toward, newell2000user}. 
InclusiveMag's methods and tools operationalize the ``one size fits all'' goal as ``one size fits two endpoints''~\cite{hamid2026inclusive}, in which each endpoint pair spans a wide range of values, and therefore a wide range of users served. 
This difference in goals from personalization means that most systems and tools in this category never monitor data about a particular user.
This avoids the ``creepiness'' that users sometimes object to when they know a system is monitoring their work (e.g.: ``I don’t like the idea of <AI product> taking definitions from my workplace. It makes me worry I'm being listened to...''~\cite{anderson2024measuringUXinclusivity}). Apply-<x>Mag follows this tradition. It reasons about facet endpoints rather than about any individual user, so it requires no data about the people who will use the interface.

\boldification{Beyond that, researchers have also studied LLM-based heuristic-violation detection more broadly}

Related to this paper is recent work exploring LLMs to automate heuristic evaluation. \citet{platt2025catching} evaluated GPT-4o across 30 websites and 850+ evaluations, finding moderate reliability in issue detection (Cohen's $\kappa=0.50$; 84\% agreement) but greater variability in severity judgments (Cohen's $\kappa=0.63$; 56\% agreement), suggesting a continued need for human oversight. Comparing GPT-4o with HCI experts, \citet{guerino2025can} found that it detected 21.2\% of expert-identified issues plus 27 additional issues, performing better on aesthetic/minimalist design and match with the real world than on flexibility, user control, and efficiency. In contrast, \citet{zhong2025synthetic} found that multimodal LLMs detected 73--77\% of usability issues, outperforming experienced UX evaluators (57--63\%), although they struggled with UI conventions and cross-screen violations. Similarly, \citet{hsueh2024applying} reported that an LLM-based tool outperformed human evaluators on user control, recognizable functions, and system status, while converging on error handling and search.

Other work has extended LLM-based evaluation to specific interfaces and evaluation settings. Duan et al.~\cite{duan2024generating} found LLMs effective at identifying visualization-heuristic violations in Figma mockups, with particular utility during early design~\cite{duan2024uicrit, kocaballi2023conversational, tabone2023using}. Lubos et al.~\cite{lubos2026recommending} combined application and task descriptions with screen recordings for MLLM-based evaluation, achieving median clarity and plausibility ratings of 4, while AIHeurEval~\cite{wang2025aiheureval} assessed consistency in color, layout, typography, and tone across UI screens. Collectively, these studies demonstrate the potential of LLMs for heuristic evaluation, but none targets a diversity dimension---the focus of InclusiveMag methods.


\boldification{One line of work pursues inclusivity by focusing on Universal Design philosophy,}

A line of work close to ours pursues inclusivity by focusing on Universal Design, fitting all users at once. The Unified User Interfaces and Design-for-All tradition builds interfaces that adapt across users, abilities, and contexts, so a single system serves the widest achievable range~\cite{stephanidis2001user, savidis2004unified}. Recent work has pushed this into automation. LLM-driven checkers such as AccessGuru detect and repair WCAG violations~\cite{fathallah2025accessguru, paterno2025llm, lopez2025turning}, steering an interface toward a universal target, adapting the design until it fits everyone. 

\boldification{Other tools target specific dimensions of inclusivity across software development, language, and education}
Alongside these universal design approaches, other tools target specific dimensions of inclusivity across software development, language, and education. \emph{Inclusivity Checker} assists developers in building more inclusive websites by evaluating accessibility and diversity-related checkpoints for different user groups \cite{pathak2023inclusivity}. At the language level, \emph{INCLUDE} promotes respectful and equitable workplace communication \cite{raja2026include}, while \citet{pomerenke2022inclusify} proposed an NLP-based approach for detecting gender-exclusive language in German and suggesting inclusive alternatives. Similarly, \citet{mihaljevic2022towards} examined tools for detecting gender bias in job advertisements, highlighting how different NLP implementations can produce substantially different assessments. \emph{WordBias} further provides an interactive approach for exploring intersectional biases across factors such as gender, race, religion, age, and socioeconomic-status in word embeddings \cite{ghai2021wordbias}. More recently, \emph{GenEDIt} extends inclusivity support to software engineering education by using an LLM-based chatbot to help educators incorporate EDI considerations into teaching materials without altering learning outcomes~\cite{arora2026genedit}.

\boldification{The InclusiveMag family pursues interface inclusivity through a different mechanism.}
%


\boldification{Mendez took the first step with a note-taking assistant; AID then automated the walkthrough itself}
\citet{mendez2018gendermag} took the first step toward automating an InclusiveMag method, building an assistant that prompts evaluators through each step of the GenderMag walkthrough and records their answers. This assistant automated the process of the walkthrough, not the detection of bugs themselves, so evaluators still had to identify each bug on their own. 
AID~\cite{chatterjee2021aid} closed that gap for one facet, using five decision rules to flag information-processing-style bugs in open-source repositories, and was later extended to all five GenderMag facets and to
courseware~\cite{chatterjee2024debugging}.  
AID's strength, however, is also its
limit: each decision rule hard-codes both the heuristic (what the facet looks
for) and the domain (how that facet's bug manifests in the artifact) into a
single hand-written rule. Because the two are fused, supporting a new facet or a new <x>Mag requires authoring a new set of decision rules.


\boldification{Geuenich et al. used LLM agents for GenderMag persona-based testing, but the system is not yet a substitute for human evaluators}
Subsequent systems broadened the target but not the generality. \citet{geuenich2026aicanseewhatyoucant} evaluated an LLM-agent system for GenderMag persona-based usability testing against traditional human-led evaluations. 
On an interface seeded with usability bugs, the agents showed limited agreement with human evaluators ($F_1 = .43$, $.45$, and $.52$): precision was high ($.83$, $1.00$, and $1.00$), but recall remained low ($.29$, $.29$, and $.35$). Thus, while their system supported multiple personas and interface types, the limited recall suggests that it would need to be paired with human evaluators continuing to do their manual evaluations.


\boldification{Anderson used LLMs to adapt code explanations to GenderMag's facet values, showing LLMs can operationalize facets beyond bug detection}
Finally, \citet{anderson2026can} extended what LLMs can do with GenderMag facets beyond detecting bugs: they studied how LLMs can adapt code explanations to people's problem-solving styles, prompting the model with GenderMag's five problem-solving-style facet values and uncovering a taxonomy of 13 distinct linguistic adaptations. 
This result 
shows that LLMs can operationalize a GenderMag facet value with enough fidelity to change their own output, not just classify someone else's. 


However, across all these lines, every InclusiveMag-family tool is hand-built for a single, method: the decision rules, the agents' personas, and the prompts each encode one method and one domain, so a new facet, <x>Mag, or combination of dimensions requires authoring them anew. No prior tool treats an <x>Mag as configuration data or automates an intersectional lens across dimensions (e.g., SES, gender, age).


\section{Apply-<x>Mag's Design Goals\draftStatus{mmb 8/16/26: D3 ! (makes my heart go pitter-patter)}} 
\label{sec:design-philosophy}
\boldification{The tool is designed for generality.  Toward this end, it aims to support ANY xMag for which there is a set of facets and a set of heuristics.}

Apply-<x>Mag aims for generality. 
As the Background section explained, <x>Mag is entirely defined by (1)~its facet set and (2)~its specialized analytic process (usually an <x>-specialized cognitive walkthrough or <x>-specialized heuristic evaluation).
The Apply-<x>Mag tool harnesses items (1) and (2) by using an <x>Mag's facets and <x>-specialized heuristics as configuration data.
Thus, the tool can support any <x>Mag for which there is a set of facets and a set of analytic heuristics.
Specifically, generality lives in configuration files, not in the tool's code, so the InclusiveMag family can grow without needing to change the Apply-<x>Mag tool for each new <x>Mag that emerges from the research community.

\boldification{Our design goals can be stated in the form of the following 4 primary use-cases: 1, 2, 3, 4.}
The tool uses an <x>Mag's configuration data to  automatically do an <x>-specialized heuristic evaluation of whatever technology product the practitioner is working with.
We state our design goals for what the tool needs to do for practitioners in the form of four primary use-cases:
\begin{enumerate}
    \item Use-case 1 (\textit{Existing} <x>Mag): A practitioner uses Apply-<x>Mag to apply an <x>Mag that existed at the time the tool was created, such as applying SESMag to their own technology product.
    \item Use-case 2 (\textit{New} <x>Mag): A practitioner uses Apply-<x>Mag to apply an <x>Mag that did not exist at the time we built Apply-<x>Mag. For example, if a group of researchers in 2028 were to create an ADHD-Mag, a practitioner could then enter the new ADHD-Mag's configuration data to apply it to their own technology product. 
    \item Use-case 3 (\textit{Extending} an <x>Mag configuration): A practitioner can extend an existing <x>Mag configuration, to either improve the tool's ability to support that <x>Mag, or to ``morph'' it into a new variant of that <x>Mag. 
    \item Use-case 4 (\textit{Intersectional  HCI}): A practitioner uses Apply-<x>Mag to apply multiple <x>Mags to find a technology product's intersectional inclusivity bugs. For example, a practitioner could apply SESMag and GenderMag to analyze socioeconomic-status and gender intersectionally as per Fallatah et al.'s method~\cite{fallatah2025intersectionalmag} (described later). 
\end{enumerate}

The next section shows how Apply-<x>Mag supports each of these use cases.
\section{How Apply-<x>Mag Works\draftStatus{top is D2.3 MMB 8/17/26}}
\label{sec:tool}

\begin{figure*}[h] 
\centering 
    \begin{minipage}{0.48\linewidth} 
        \centering 
        \includegraphics[width=.9\linewidth]{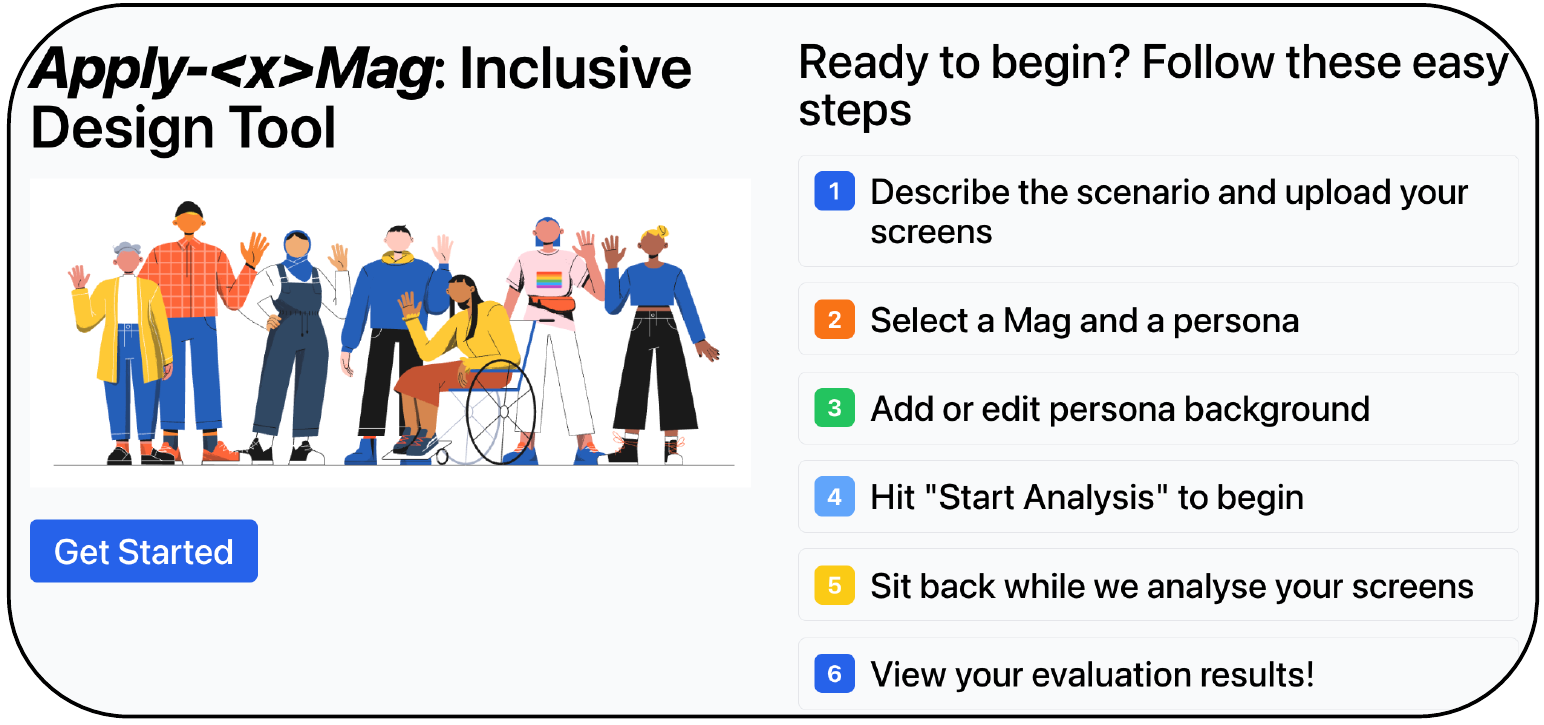}
        \captionof{figure}{
        The tool's landing page tells the practitioner what to do, so ... (cont. on Figure~\ref{fig:sample-input-screenshot}).}
        \label{fig:landing-page}
        \Description{A screenshot of the tool’s landing page titled “Apply-<x>Mag: Inclusive Design Tool.” Below the title on the left is an image depicting a diverse group of people raising their hands, followed by a blue button labeled “Get Started.” On the right, a header reads “Ready to begin? Follow these easy steps.” Below is a list of 6 steps. 1: Describe the scenario and upload your screens. 2: Select a Mag and a persona. 3: Add or edit persona background. 4: Hit “Start Analysis” to begin. 5: Sit back while we analyse your screens. 6: View your evaluation results!}
    \end{minipage} 
    \hfill
    \begin{minipage}{0.48\linewidth} 
        \centering
        \includegraphics[width=0.9\linewidth]{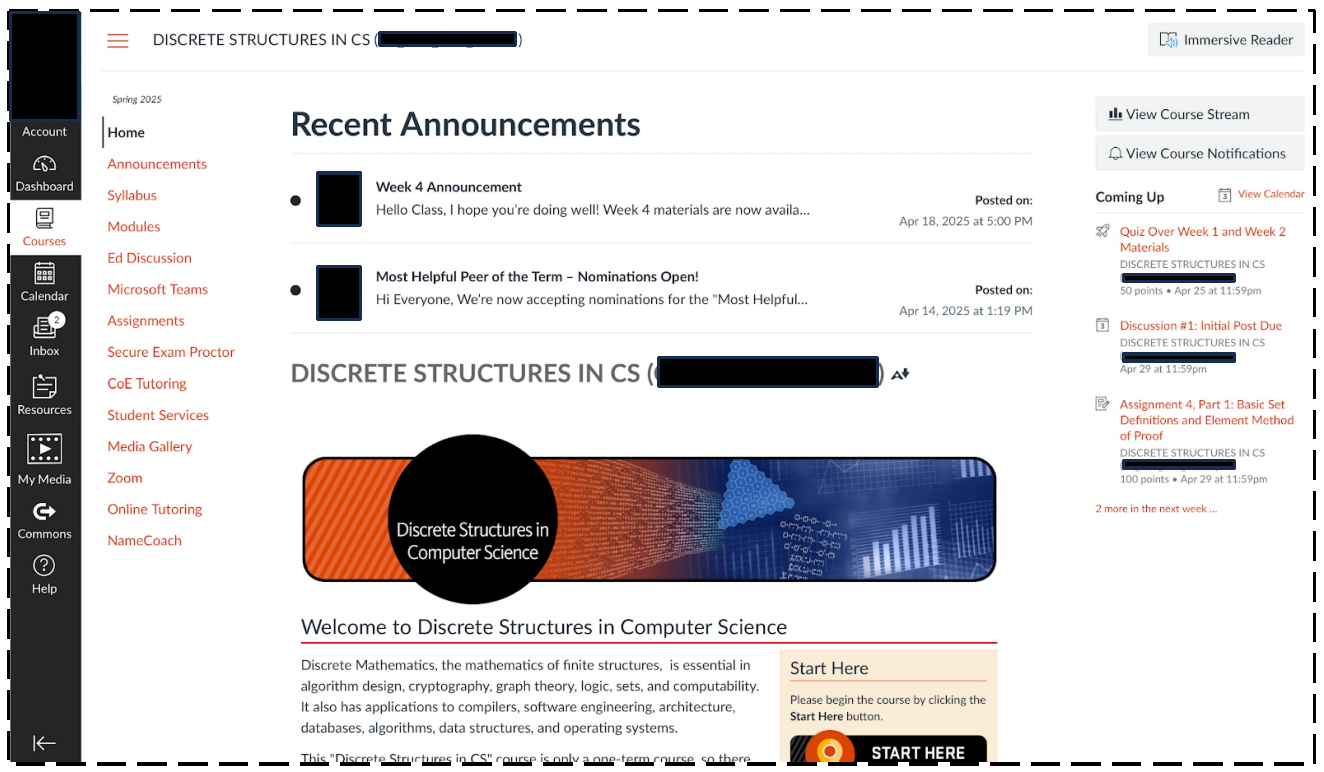}
        \caption{
        (cont. from Figure~\ref{fig:landing-page}) ... so the practitioner begins by entering a textual scenario description (not shown) and their screens. Here they are working on web-based ``courseware'' for CS-Math 
course, so this is one of the screens they upload. (Larger image in Supplemental Docs.)\\}
        \label{fig:sample-input-screenshot}
        \Description{A screenshot of a learning management system, showing the page for the course Discrete Structures in Computer Science. The page features recent announcements, a large banner image, a welcome paragraph, and button to the side labeled “START HERE.”}
    \end{minipage} 

    \begin{minipage} {0.9\linewidth}
        \centering
        \includegraphics[width=1\linewidth]{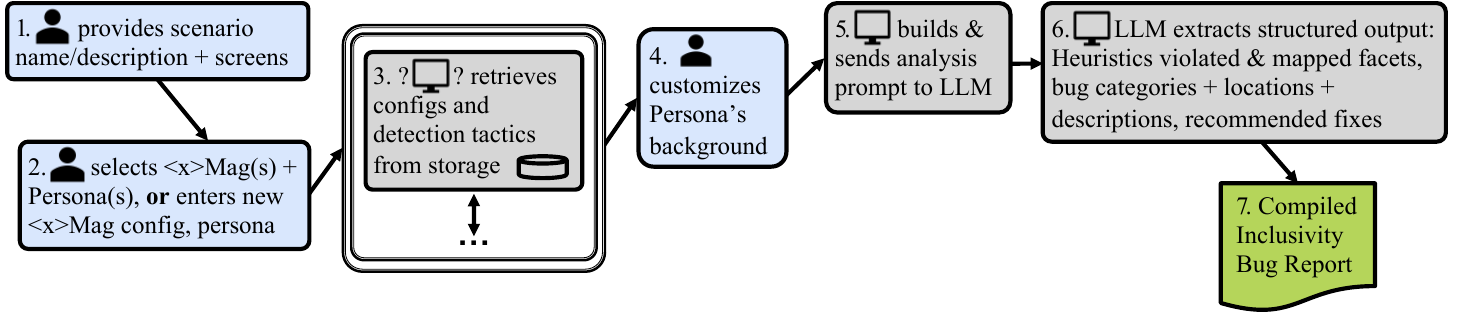}
        \caption{How Apply-<x>Mag works.  Arrows show flow of data, and numbering indicates the sequence in which events take place. Double box (Box~3) is expanded in detail later in Figure~\ref{fig:step3archi}.
        \textcolor{userBlue}{Blue} boxes: practitioner inputs; \textcolor{gray}{gray} boxes: tool's actions; \textcolor{outputGreen}{green} box: produced by the tool. \added[id={FAM}]{\textbf{?<symbol>?}}: conditional depending on what is already stored.}
        \label{fig:tool-archi}
        \Description{Flowchart with seven numbered steps. Beginning with Step-1, they read: (person icon) provides scenario name/description + screens. Step-1 points to Step-2: (person icon) selects <x>Mag(s) + Persona(s), or enters new <x>Mag config, persona. Step-2 points to Step-3 inside a bounded box: (computer icon surrounded by question marks) retrieves configs and detection tactics from storage (database icon). Within the bounded box, a bidirectional arrow points between Step-3 and an ellipsis. Another arrow points from Step-3 out of the bounding box to Step 4: (person icon) customizes Persona’s background. Step 4 points to Step 5: (computer icon) builds and sends analysis prompt to LLM. Step 5 points to Step 6: (computer icon) LLM extracts structured output: Heuristics violated and mapped facets, bug categories + locations + descriptions, recommended fixes. Step 6 points to Step 7, which is the endpoint: Compiled Inclusivity Bug Report.}
    \end{minipage} 
\end{figure*}

\boldification{The landing page tells practitioner what to do.}
The tool begins with a landing page (Figure~\ref{fig:landing-page}), which tells the practitioner the steps to take.
The way the tool reasons internally, given the practitioner's inputs and actions, is depicted by Figure~\ref{fig:tool-archi}. 
As  Figure~\ref{fig:tool-archi} shows, the tool retrieves the configuration files that match the practitioners' inputs, and after any further customization the practitioner does, sends a prompt to the LLM, which returns the data and formats it into an inclusivity bug report.

\boldification{The practitioner begins by providing the tool with a scenario they want to work on: a text description and a sequence of screens they propose should accomplish that scenario.}
For all four of the use-cases, the practitioner begins by providing the tool with a scenario they want to work on: a text description and a sequence of screens they propose would accomplish that scenario (Figure~\ref{fig:tool-archi}'s first blue box, top left).
Practitioners provide these screens via scanned-in sketches or screenshots (pdf, png, etc.), which enables iterating on even early designs and mockups.%
\footnote{Mature implemented products can also be entered this way, but future implementations could facilitate further by also processing other formats, such as HTML files as in Chatterjee et al.'s GenderMag tool for analyzing courseware~\cite{chatterjee2022icer}.
} 
E.g., if a practitioner like an instructional designer is working on some online ``courseware'', one of their screens might look like Figure~\ref{fig:sample-input-screenshot}.  
What the practitioner does next varies by use-case, as detailed next.




\subsection{Use-case 1: Practitioner uses Apply-<x>Mag for a pre-existing <x>Mag\draftStatus{D3 MMB 8/26/26}}
\label{subsec:Use-case1}
\label{subsubsec:SESMag}

\boldification{Practitioners picks SESMag, which is...}

\begin{figure*} [!t]
    \centering
    \includegraphics[width=0.9\linewidth]{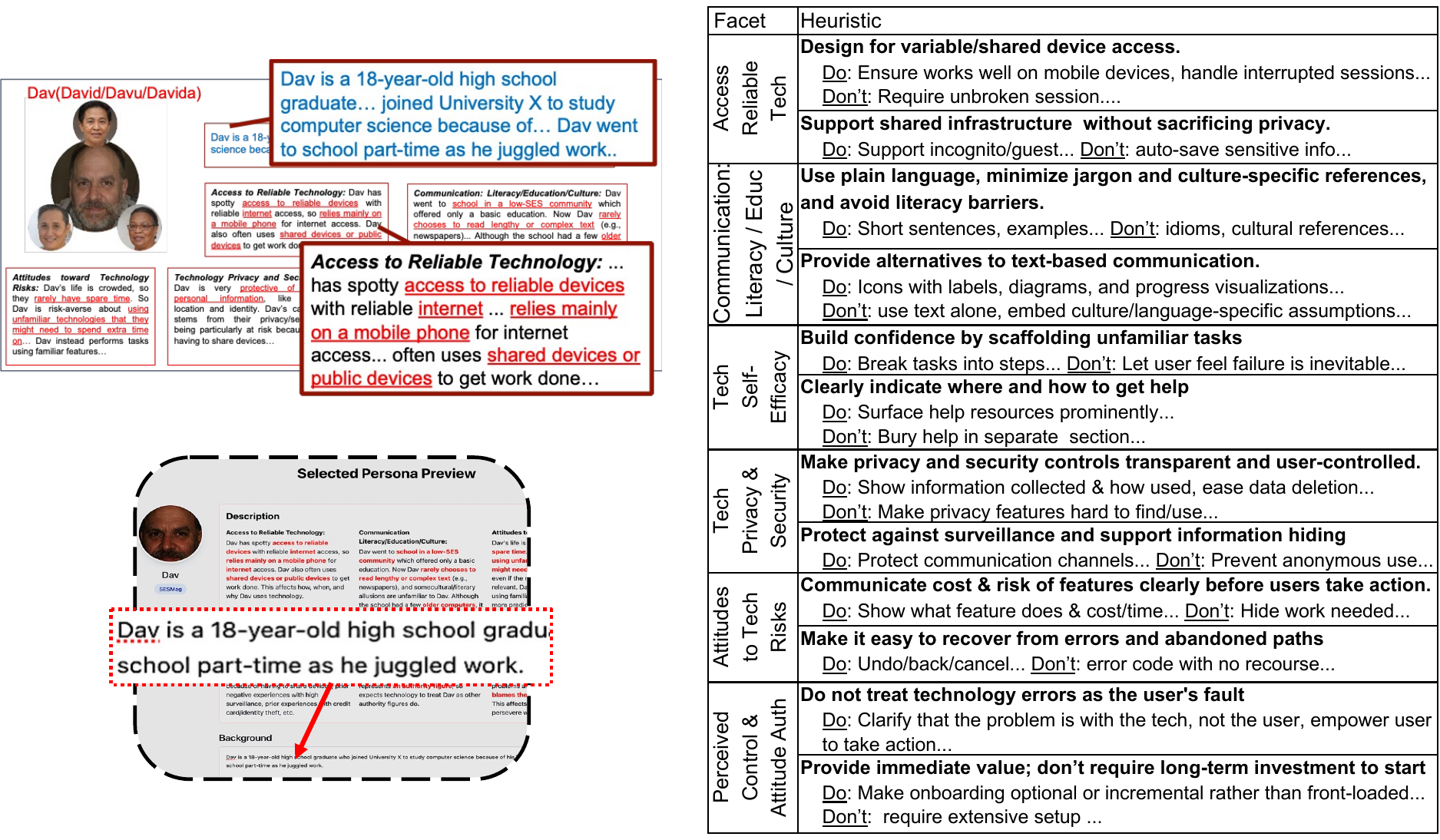}
    \caption{
    The tool's configuration+background data: (Top left): The tool already knows the Dav persona, shown here as used by our participants manually with portions magnified. See Supp. Docs for full Dav persona. The blue text (``Dav is a 18...'') is customizable. (Bottom left): The practitioner sees a preview of the Dav persona, and optionally customizes Dav's background (``Dav is a 18...''). (Right:) The tool also already knows each SESMag facet type's heuristics (do's and don'ts abbreviated here; unabbreviated in Supp. Docs.), that were derived from the SESMag foundations paper~\cite{burnett2024SESMag}.}
    \label{fig:SESMag-Persona+Heuristics}
    \label{fig:dav}
    \label{fig:persona-customize}
    \label{fig:sesmag_heuristics_table}
    \Description{(Top left) The Dav persona, with an image of several different faces that represent Dav, as well as their facets. Dav’s background and Access to Reliable Technology facet are enlarged in callouts to make them readable. The background reads: “Dav is an 18-year old high school graduate… joined University X to study computer science because of… Dav went to school part-time as he juggled work.” The Access to Reliable Technology facet reads: “...has spotty access to reliable devices with reliable internet… relies mainly on a mobile phone for internet access… often uses shared devices or public devices to get work done…”\\Bottom left) A screenshot of the tool showing a preview of the Dav persona selected by the practitioner. There is a picture and description of the facets, and a blown-up callout with an arrow pointing to Dav’s background, which reads: “Dav is a 18-year old high school gradu…” <cut off>  “...school part-time as he juggled work.”}
\end{figure*}

In Use-case 1, a practitioner has decided to evaluate their technology's inclusiveness using one of the <x>Mags.
During start-up processing, the tool pulled the names of known <x>Mags and personas from its storage; it now offers the known <x>Mags for selection.
In this example, they choose SESMag to evaluate the scenario they entered above, as per Figure~\ref{fig:tool-archi}'s Box~2.
SESMag (Socioeconomic Inclusiveness Magnifier) is a relatively new member of the InclusiveMag family, for use in iteratively designing and/or evaluating a tech product's inclusivity for individuals across the socioeconomic spectrum \cite{burnett2024SESMag, chikezie2025SESmagSurvey, busteed2026}.
As with all <x>Mag methods, at the core of SESMag are its facet types---ranges of individual traits relevant to problem-solving whose values differ statistically across SES strata.
Figure~\ref{fig:sesmag_heuristics_table} (right) lists the six SESMag facet types. 

\boldification{next the practitioner chooses Dav.}

Since SESMag is one of the configurations that the tool already knows, after the practitioner selects SESMag from the menu, they can select one of the personas that has been configured for that Mag.
The processes for creating a new <x>Mag and a new persona are described later in Use-cases~2 and~3, respectively.


In this example, the practitioner completes Figure~\ref{fig:tool-archi}'s Box~2 by choosing Dav (Figure~\ref{fig:dav}'s top left).
As Section~\ref{subsec:background} explained earlier, Dav is entirely defined by the six endpoint values on the lower-SES end of the SESMag facet type ranges, one of which is blown up for readability in Figure~\ref{fig:dav}. 
This triggers the tool's backend as follows (\colorbox{gray!25}{gray text} in the text body and in Figure~\ref{fig:tool-archi}).
\vspace{1mm}

\boldification{The Mag and Persona tell the backend which configuration data to retrieve from its collection of configurations. } 

\mybox{
\textit{The configuration data}:
As soon as the practitioner has entered the Mag and the persona, the tool's backend retrieves the configuration data for that <x>Mag (SESMag) and that persona (Dav), as per Figure~\ref{fig:tool-archi}'s Box~3.
For SESMag's Dav, the configuration consists of Dav's facet values, the SESMag heuristics \cite{burnett2024SESMag} shown in Figure~\ref{fig:sesmag_heuristics_table} for each facet type, and  (optional) examples of bugs, solutions, and bug categories for each heuristic.
(SESMag's complete configuration is in the Supp. Docs.) Since this section's use-case applies an <x>Mag whose configuration the tool previously stored, practitioners need not concern themselves with configurations.
} 

\boldification{The practitioner optionally can customize the background data, so they do. That's it -- practitioner is done.}

At this point, the tool responds to the practitioner with a preview of the persona Dav. 
Practitioners can optionally customize Dav with suitable background for \textit{their} product (Figure~\ref{fig:persona-customize}, bottom left).
This completes Figure~\ref{fig:tool-archi}'s Box~4.
There is nothing else the practitioner needs to enter---\colorbox{gray!25}{the tool} takes over from here.

\vspace{1mm}

\boldification{Based on the configuration data, the backend do its own heuristic evaluation (how -- we'll discuss later)}

\mybox{
 \textit{The inclusivity bug report}:
To handle Figure~\ref{fig:tool-archi}'s Boxes 4--7, the tool turns the combination of the scenario and screens with the configuration data into one to two LLM prompts, using mechanisms we detail in the next section.
The LLM's response to the last prompt generates  an inclusivity bug report detailing the inclusivity bugs it found on each uploaded screen. 
} 

\boldification{The tool follows why/where/fix.}

\mybox{
The tool generates the report as per Guizani et al.'s \textit{Why/Where/Fix} inclusivity debugging approach~\cite{guizani2022whywherefix}.
In Guizani et al., Why/Where/Fix was described as a manual process: human evaluators use the persona's facet values to figure out that an inclusivity bug existed, \textit{why} it exists, \textit{where} the bug is---i.e., in what elements of the information architecture (IA) elements~\cite{rosenfeld1998informationArch} (e.g., links, screens, labels, terms or phrases, relationships, between and among groups)---and what kind of \textit{fix} the why+where together imply.
Our tool automatically performs Why/Where/Fix.
Its \textit{why} is based on the facets and heuristics, its \textit{where} refers to IA elements, and its \textit{fix} recommendations come from not only \textit{why}+\textit{where} but also from worked examples if provided with the configuration.
} 

\boldification{The practitioner sees the why/where/fix inclusivity report for all images}

\begin{figure*}[tbp]
    \centering 
    \includegraphics[width=0.9\linewidth]{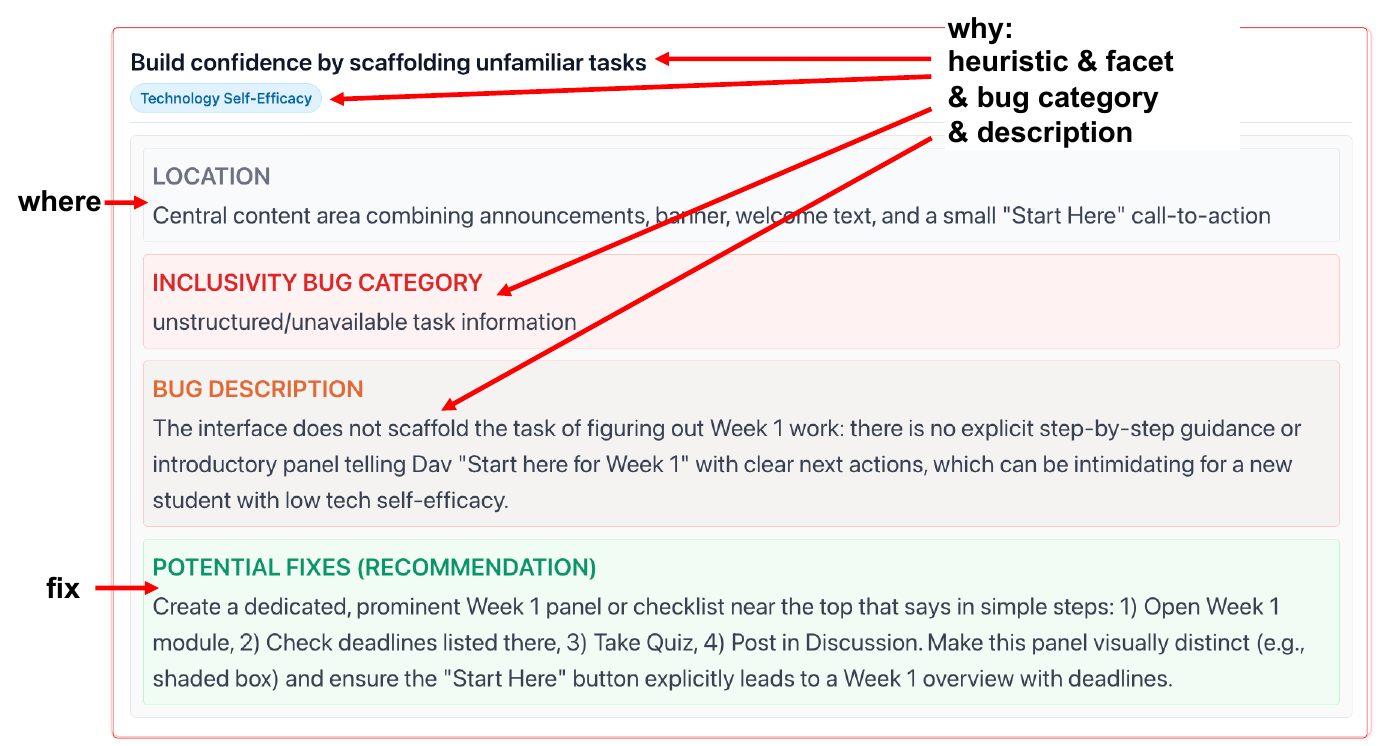}
    \captionof{figure}{The tool produces an inclusivity bug report showing the why+where+fix as per Guizani et al.'s Why/Where/Fix inclusivity debugging approach~\cite{guizani2022whywherefix}. Accompanying this SESMag bug report is the CS-Math courseware screen with the bug, which in this example is the same as Figure~\ref{fig:sample-input-screenshot}.}
    \label{fig:bug_report}
    \Description{A screenshot of the tool-generated inclusivity bug report. The title reads: Build confidence by scaffolding unfamiliar tasks. Just below the title is a blue tag labeled with the facet: Technology Self-Efficacy. Next are the location, inclusivity bug category, bug description, and potential fixes. LOCATION: Central content area combining announcements, banner, welcome text, and a small “Start Here” call to action. INCLUSIVITY BUG CATEGORY: unstructured/unavailable task information. BUG DESCRIPTION: The interface does not scaffold the task of figuring out Week 1 work: there is no explicit step-by-step guidance or introductory panel telling Dav “Start here for Week 1” with clear next actions, which can be intimidating for a new student with low tech self-efficacy. POTENTIAL FIXES (RECOMMENDATION): Create a dedicated, prominent Week 1 panel or checklist near the top that says in simple steps: 1) Open Week 1 module, 2) Check deadlines listed there, 3) Take Quiz, 4) Post in Discussion. Make this panel visually distinct (e.g., shaded box) and ensure the “Start Here” button explicitly leads to a Week 1 overview with deadlines. In the top right corner, an annotation with arrows shows how the report follows Guizani et al.’s Why/Where/Fix approach. “Why” points to the title. “Where” points to the LOCATION box. “Heuristic & Facet” points to the facet tag under the title. “Bug category” points to the INCLUSIVITY BUG CATEGORY box. “Description” points to the BUG DESCRIPTION box. “Fix” points to POTENTIAL FIXES (RECOMMENDATION) box.}
\end{figure*}

As a result, the practitioner receives a Why/Where/Fix inclusivity bug report like the one in Figure~\ref{fig:bug_report} for each screen, along with the screenshot (here, same as Figure~\ref{fig:sample-input-screenshot}) with a box drawn around the IA element(s) containing the bug.
As per the definition of inclusivity bugs in Section~\ref{sec:intro}, these bugs are usability bugs found via Dav's facet values, which are thus disproportionately likely to affect someone with Dav's facet values.

\boldification{Beyond Dav to the whole range.}

But what about inclusivity bugs affecting people whose facet values aren't like Dav's?
Dav represents the SESMag facets' lower-SES endpoints only.
To harness the full range of SESMag facet values, the practitioner will need to run the tool once more---but only once.
Specifically, the practitioner simply repeats the process with Fee, who represents the higher-SES facet endpoints.
They reuse the same scenario and screens, and receive the set of inclusivity bugs disproportionately affecting Fee.

\boldification{Endpoints are powerful!}

This is where the expressive power of facet endpoints becomes clear.
As per InclusiveMag family practices (Section~\ref{subsec:background}), each fix the practitioner makes to an inclusivity bug will need to work for both Dav and Fee \textit{simultaneously}---as opposed to building one feature set for Dav and a different feature set for Fee.
As has been empirically verified for various <x>Mags (e.g.,~\cite{agarwal2023sesmag,guizani2022whywherefix,vorvoreanu2019genderMag,xiao2026elderlyHealthmag}), 
this simultaneousness of endpoint coverage ensures inclusivity of entire ranges, including arbitrary combinations of facet values within these ranges.


\subsection{Use-case 2: Practitioner uses Apply-<x>Mag on a new <x>Mag\draftStatus{MMB 8/24: d2.5}}
\label{subsec:Use-case2}

\boldification{When researchers come up with new Mag, tool supports adding new Mag for practioners to use.}

Suppose a research group in Denmark creates a new <x>Mag named ADHD-Mag and publishes it in CHI'28, including ADHD-Mag's facet set and heuristics for an ADHD-specialized heuristic evaluation.
Upon reading the CHI paper, suppose a practitioner in Brazil has a ``this is just what I need!'' insight about ADHD-Mag's potential value for one of their technology products.
The Brazilian practitioner can act upon this insight by entering a configuration using the ADHD-Mag CHI'28 paper.

\boldification{To enter the ADHD-Mag configuration..}

The entry process requires only ordinary text entry plus a spreadsheet file upload:
The practitioner textually enters the ADHD-Mag description, facet types and heuristics, then uploads a spreadsheet (current accepted format: csv) with the heuristics and facet types, (e.g., as in Figure~\ref{fig:sesmag_heuristics_table}).
To complete the configuration, they enter at least two ADHD-Mag personas (one for each end of the facet types' endpoint values) with URLs for photos, which look similar to the Dav persona in Figure~\ref{fig:dav}'s lower left.
This completes the minimum configuration needed, so the practitioner can now apply ADHD-Mag to their products as in Use-case~1 (Section~\ref{subsec:Use-case1}).

\boldification{Here's how the tool uses the configuration data to figures out guidance to send to the LLM for how to detect such bugs.  If optional info has been entered, it uses that to help; otherwise it does so using only the facets, background, and heuristics.}

\begin{figure*} [h]
    \centering
    \includegraphics[width=0.9\linewidth]{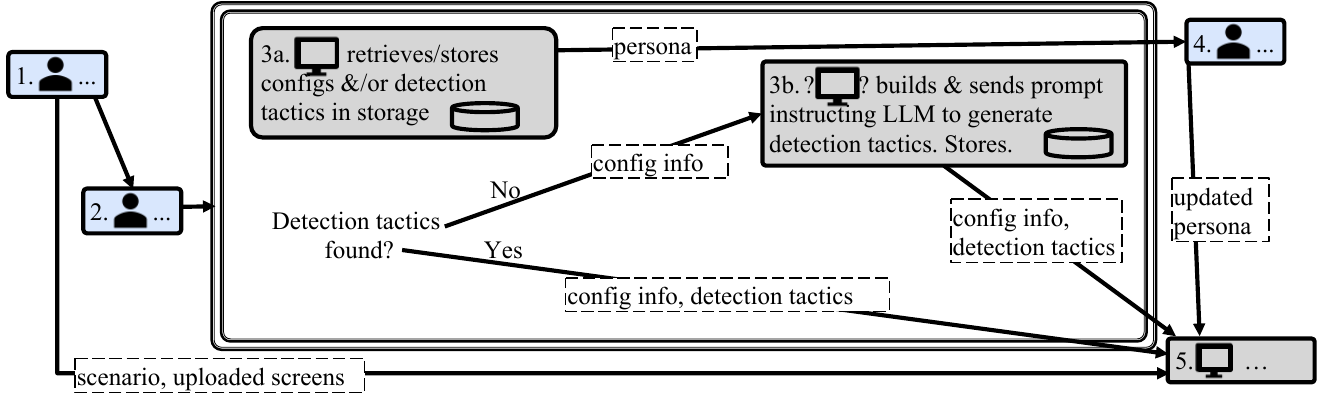}
    \caption{Figure~\ref{fig:tool-archi}'s Box~3 expanded. (Colors, symbols same as Figure~\ref{fig:tool-archi}.) Dashed text labels data flowing. In essence, if the tool has stored an <x>Mag's detection tactics from a prior run, it omits Box~3b; otherwise Box~3b prompts an LLM to generate the detection tactics and then stores them. }
    \label{fig:step3archi}
    \Description{Flowchart with five numbered steps. From Step-1 (person icon) one path leads to Step-2 (person icon). A second path runs from Step-1 directly to Step 5, labeled "scenario, uploaded screens.” Step-2 enters a large bounded region containing steps 3a and 3b and a decision point. Inside the bounded region: Step-3a (computer icon, with an attached database icon) reads "retrieves/stores configs and/or detection tactics in storage." From 3a, an arrow labeled "persona" points right, out of the region, to Step 4. Below 3a is a decision point labeled "Detection tactics found?" with two branches. The "No" branch, labeled "config info," points to Step-3b. The "Yes" branch, labeled "config info, detection tactics," points down and out of the region to Step 5. Step-3b (computer icon, surrounded by question marks) reads "builds and sends prompt instructing LLM to generate detection tactics. Stores (database icon)." From 3b, an arrow labeled "config info, detection tactics" points down to Step 5. Step 4 (person icon), reached from 3a via the "persona" arrow, sends an arrow labeled "updated persona" down to Step 5. Step 5 (computer icon) is the endpoint, receiving four incoming arrows: "scenario, uploaded screens" from Steps 1/2, "config info, detection tactics" from the decision point's "Yes" branch, "config info, detection tactics" from Step-3b, and "updated persona" from Step 4.}
\end{figure*}

\mybox{
 \textit{How the tool generates Detection Tactics and the Why/Where/Fix bug report:} 
Given the minimal configuration entered by the Brazilian practitioner, Apply-<x>Mag now looks up ADHD-Mag's configuration in global storage's hashtable and does not find one (Figure~\ref{fig:step3archi}'s Box~3a). 
Since ADHD-Mag is new, Apply-<x>Mag next prompts the LLM to produce detection tactics for each heuristic in ADHD-Mag's configuration (Box~3b). 
Here the practitioner did not provide examples, so it is a zero-shot prompt.
Apply-<x>Mag then stores the resulting tactics.
The now-stored ADHD-Mag configuration and detection tactics let future uses of ADHD-Mag skip Box~3b's LLM call, reducing resource consumption and improving efficiency.
} 

\mybox{
Switching to SESMag for an example from our logs, sending such a prompt to the LLM for the heuristic ``Design for variable and shared device access'' generated several detection tactics, one of which was ``Check if practitioners can save partial work and return later (drafts, autosave, or a clear `Resume entry).''
As shown in Figure \ref{fig:step3archi}, the tool then packages up the generated detection tactics, persona facet values, customized persona background, scenario, and product screenshot(s) into a multimodal prompt to the LLM (zero-shot), instructing it to find inclusivity bugs and draw a bounding box around each bug's location on the screen.
}

\mybox{
To reduce false positives, the prompt also includes this instruction: \textit{``Only select a heuristic violation where you are HIGHLY CONFIDENT ($\approx$80\% or higher confidence) of having observable violations supported by visible UI evidence in the screenshot. If confidence is below that threshold, DO NOT report that.''} 
(Examples of the prompts are in the Supp. Docs.)
Using these instructions, the LLM returns completed Why/Where/Fix bug report(s), such as in Figure~\ref{fig:bug_report}, along with the screen(s) updated with bounding box(es).
} 

\subsection{Use-case 3: Practitioner extends an <x>-Mag configuration\draftStatus{MMB: 2.5 8/23/26}}
\label{subsec:Use-case3}

Practitioners can extend an <x>Mag in several ways.

\boldification{Way 1: When they run the tool, they can feed part of the results back into an expanded configuration as follows...}

One way is to add worked examples of prior ADHD-inclusivity bugs to the configuration spreadsheet.
If they update the configuration in this way, the tool will remember these worked examples for future uses.
For example, after applying the new ADHD-Mag to their own products, the Brazilian practitioner can optionally harvest the bugs and fixes to update the ADHD-Mag configuration spreadsheet with these worked examples. 
Another source of worked examples is prior literature. For example, the Brazilian practitioner could add the following known ADHD-inclusivity bug and solution from the work of \citet{lim2026supportADHD} (excerpted for brevity): 

\borderedbox{
Violation: Tutorial too wordy, too rigid... <user> overwhelmed.\newline
Solution:  Break into small, bitesized steps that feel doable.
}




\mybox{
\textit{How Apply-<x>Mag generates Detection Tactics now:}
When configuration data contains worked examples as above, the tool can harness the examples by sending a few-shot prompt instead of a zero-shot prompt (Figure~\ref{fig:tool-archi}'s Box 5). 
Otherwise, the tool operates the same as before.
} 

\boldification{Way 2:  add a new persona}

Extending an <x>Mag in other ways is possible too. 
For example, the Brazilian practitioner may want to add a third persona, this one with mixed facet values. 
They can simply create a new persona with the same facet \textit{types} as the other ADHD-Mag personas (e.g., attention span) but with different \textit{values} (e.g., \textit{medium} attention span \textit{but easily distractible}).

\boldification{Way 3: or add a new facet type w heuristics -> Brazilian-ADHD-Mag}

Finally, suppose the Brazilian practitioner realizes that the Danish researchers' ADHD-Mag lacks a facet ~\textit{type} important in Brazil, such as access to medical care for ADHD~\cite{ortega2020adhd-brazil}.  
The practitioner can create a new Brazilian-ADHD-Mag by 
adding the new facet type (access to ADHD medical care) and  heuristics to ADHD-Mag's configuration spreadsheet, and update the personas with facet values for that new facet type.  
They save it with a new <x>Mag name such as Brazilian-ADHD-Mag, and proceed from there as in Use-case~2.


\newcommand{\rowlabel}[2]{\rotatebox[origin=c]{90}{\shortstack{\emph{#1}\\\emph{#2}}}}
\newcommand{\tbullets}[1]{%
  \begingroup\linespread{0.92}\selectfont
  \begin{itemize}[nosep, topsep=1pt, partopsep=0pt, parsep=0pt, itemsep=0pt,
                  leftmargin=1.1em, label=\textbullet]#1\end{itemize}\endgroup}

\begin{table}[ht]
\centering
\linespread{0.94}\selectfont
\setlength{\tabcolsep}{2pt}
\renewcommand{\arraystretch}{0.95}
\setlength{\aboverulesep}{1pt}
\setlength{\belowrulesep}{1pt}
\caption{
What <x>Mags vs.\ quantitative empirical studies reason about.
These differences in reasoning explain why one-dimensional <x>Mags' inclusivity bugs can be union'd together to find intersectional sets of bugs, whereas one-dimensional quantitative empirical study results cannot.}
\label{tab:xmagVSempirical}
\begin{tabular}{@{} >{\centering\arraybackslash}m{2.8em}
                    >{\raggedright\arraybackslash}p{0.52\linewidth}
                    >{\raggedright\arraybackslash}p{0.34\linewidth} @{}}
\toprule
 & \textbf{<x>Mags (empirical)} & \textbf{Empirical studies (quantitative)} \\
\midrule
\multirow{10}{*}{\rowlabel{Reasoning}{about\ldots}}
  & People's personal traits (facet values). \textit{Examples}:
    \tbullets{%
      \item Has access to reliable tech
      \item Was educated at a highly-resourced school
      \item Has high computer self-efficacy\ldots}
  & People's identities. \textit{Examples}:
    \tbullets{%
      \item Has high SES} \\
\cmidrule(l{\tabcolsep}){2-3}
  & \textit{Examples (cont)}:
    \tbullets{%
      \item Has a short attention span\ldots}
  & \textit{Examples (cont)}:
    \tbullets{%
      \item Has ADHD} \\
\midrule
\multirow{13}{*}{\rowlabel{Reasoning}{covers\ldots}}
  & Range (all possible values between the endpoints) of every facet type. \textit{Examples}:
    \tbullets{%
      \item Access range: from low to high access to reliable tech
      \item Education range: from under-resourced to highly-resourced schools
      \item Attention span range: from short to long attention spans\ldots}
  & Population sample of individual data points collected. \textit{Examples}:
    \tbullets{%
      \item 10 low-SES people with ADHD +\newline 
      500 high-SES people with ADHD +\newline
      600 low-SES people without ADHD +\newline
      1000 high-SES people without ADHD} \\
\bottomrule
\end{tabular}
\end{table}


\begin{figure*}[h]
    \centering
    \includegraphics[width=0.9\textwidth]{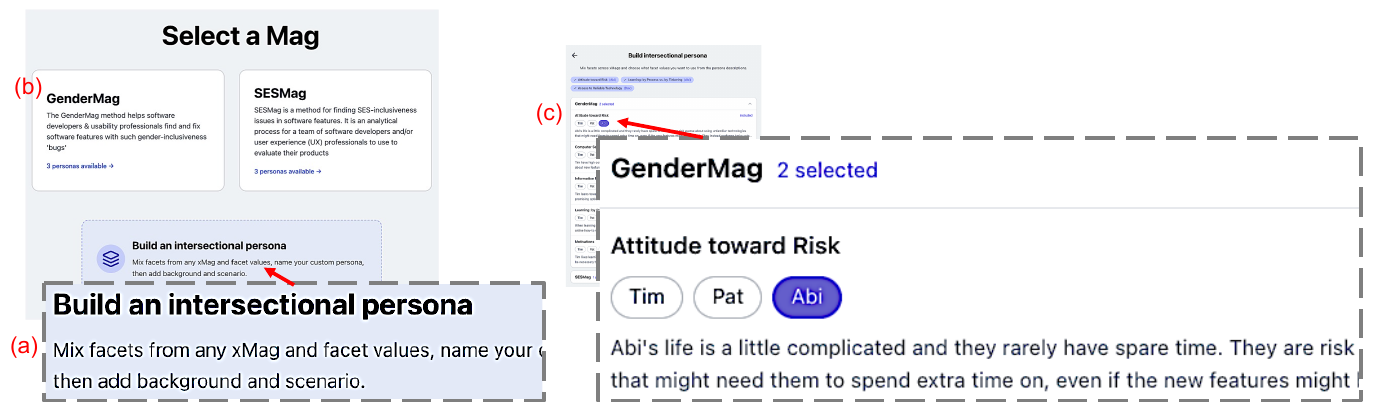}
    \caption{
    The practitioner is creating an intersectional GenderSESMag by combining two existing <x>Mags, GenderMag and SESMag. (a) As the instructions say, the practitioner has only to put the intersectional personas together as desired. (b) They begin by selecting GenderMag, then (c) select the desired ``Abi'' facet values, which they will eventually combine with SESMag's ``Dav'' facet values. The resulting Abi+Dav persona will represent the intersectional population similar to both Abi (whose facet values are predominantly those of women) and Dav (facet values predominantly those of lower-SES individuals).}
    \label{fig:intersectional}
    \Description{Left: A screenshot of the tool with the header “Select a Mag.” Below are two boxes, one labeled GenderMag (with an annotated red letter “b”). Next to it is another box labeled SESMag. There is also a button labeled “Build an intersectional persona” with a stack icon. A callout pointing to the button (with annotated red letter “a”) reads: Mix facets from any xMag and facet values, name your… <cut off> …then add background and scenario. Right: a screenshot of the tool titled “Build intersectional persona.” A callout points to the image with the header “GenderMag (2 selected).” Below, the facet Attitude toward Risk and three buttons representing the personas Tim, Pat and Abi, with Abi’s button selected, and the text: Abi’s life is a little complicated and they rarely have spare time. They are risk… <cut off> that might need them to spend extra time on, even if the new features might… <cut off>.
}
\end{figure*}

\subsection{Use-case 4: Practitioner uses Apply-<x>Mag for intersectional HCI\draftStatus{mmb 8/30 d2.5}}
\label{subsec:Use-case4}


\boldification{As many have pointed out, a one-dimensional approach to inclusive design may not be enough. So the tool needs to support intersectional HCI.}

As many researchers point out~\cite{andalibi2022LGBTQ, buolamwini2018gender, crenshaw1991intersectional, erete2018intersectional, rankin2020intersectional, 
schlesinger2017intersectional, wisniewski2018intersectionality}, one-dimensional approaches to inclusive design have limitations.
Intersectional HCI
\footnote{As in Rankin et al. and Fallatah et al.~\cite{fallatah2025intersectionalmag, rankin2019intersectionalHCI}, we distinguish intersectional HCI from intersectionality. Intersectional HCI \textit{applies} aspects of intersectionality to technology. 
}
~recognizes that humans' interconnected social identities shape their experiences with technology, so
considers technology products' inclusivity for multiple diversity dimensions at once~\cite{fallatah2025intersectionalmag}, such as SES$\times$ADHD.
For example, an intersectional SES$\times$ADHD analysis would consider how user experiences differ between four quadrants: low-SES users with ADHD, high-SES with ADHD, low-SES without ADHD, and high-SES without ADHD. 

\boldification{Apply-<x>Mag supports intersectional HCI in **five** ways}

The Apply-<x>Mag tool supports intersectional HCI in five ways: (1)~apply an existing intersectional <x>Mag using  Use-case~1's process;
\footnote{  
Intersectional <x>Mags to date are Elderly HealthMag~\cite{xiao2026elderlyHealthmag} and the intersectional <x>Mag in \citet{fallatah2025intersectionalmag}.}
(2)~apply a new intersectional <x>Mag by entering it as per Use-case~2;
(3)~``morph'' an existing single-dimensional <x>Mag into a (new) intersectional <x>Mag by adding new facets and heuristics, as per Use-case~3;  
(4)~use multiple one-dimensional <x>Mags serially and take the union of both methods' resulting inclusivity bugs; and 
(5)~compose two or more <x>Mags into a new <x>Mag and  apply the new <x>Mag as usual. 
The first three are obviously supported by Use-cases~1--3, but the last two warrant further explanation.

\boldification{In approach 4, you run xMag and then xMag. But how can this work?}

The serial approach to intersectional HCI (item (4)) rests on the premise that practitioners can run an intersectional analysis by simply running two different <x>Mags (e.g., SESMag and then ADHD-Mag) and then taking a union of the inclusivity bugs they find.
Yet, empirical intersectionality research shows the opposite (e.g.,~\cite{Buolamwini2019WrittenTestimony}).
This raises the question: how could both be true?

\boldification{It does, because empirical <> analytical. Fallatah et al. showed  theoretically and empirically that this works in combining <x>Mag results, even though it doesn't work in combining empirical results.}

The answer lies in the reasoning bases behind these methods, summarized in Table~\ref{tab:xmagVSempirical}.
As \citet{fallatah2025intersectionalmag} explain, statistically analyzing empirical data one identity at a time overlooks intersectional groups who happen to be underrepresented in the data (e.g.,~\cite{buolamwini2017gender, Buolamwini2019WrittenTestimony}).
Achieving equal sample sizes across intersectional identity groups is a challenging problem in quantitative empirical studies (e.g.,~\cite{ahmed2018trans, mason2025diverseParticipants, sarkar2024diverseParticipants, liang2021researchMarginalized}).
In Table~\ref{tab:xmagVSempirical}'s hypothetical example of an SES$\times$ADHD investigation, when analyzing low- vs. high-SES (alone), the 10 low-SES people with ADHD would make no statistical difference if combined with the 600 low-SES people without ADHD.
Similarly, when analyzing with-ADHD vs. without-ADHD (alone), the 10 low-SES people with ADHD would make no statistical difference when combined with the 500 high-SES people with ADHD.
The <x>Mags avoid this problem because they avoid reasoning about people's identities, do not reason with population data samples, and do not reason statistically. This union validity was established by Fallatah et al.~\cite{fallatah2025intersectionalmag}, and we return to it in Section~\ref{subsec:results-Use-case4}.
        
\boldification{The compositional approach is functionally equiv to Approach 4, but takes care of the union'ing automatically.}

The compositional approach runs multiple <x>Mags together. 
It looks different on the front end (Figure~\ref{fig:intersectional}), in that the practitioner says which <x>Mags and facets they want to combine for an intersectional analysis.
The compositional approach produces the same bugs as the serial approach, but the practitioner's work is different.
Up-front, the compositional approach requires the practitioner to set up intersectional personas.
But this up-front work leads to less work per run than the serial approach because the tool will now automatically synthesize all the inclusivity bugs into one set of inclusivity bug reports per screen.

\begin{table*}[h]
\centering
\caption{
Overview of human-versus-tool evaluations, comparing manual SESMag and GenderMag results with tool results and participant validation. (Note: Some screens for which humans found no bugs were missing from our field data, so are not included here.)
}
\label{tab:studyoverview}
\resizebox{0.8\linewidth}{!}{%
\setlength{\tabcolsep}{2pt}
\begin{tabular}{@{}cllcll c l@{}}
\toprule
\textbf{Context} & \textbf{Type} & \textbf{Product} & \textbf{\# Screens} & \textbf{Method} 
    & \makecell[l]{\textbf{Manual}\\\textbf{Eval.\ by\ldots}}
    & \makecell[c]{\textbf{Tool}\\\textbf{Eval.}}
    & \makecell[l]{\textbf{Participant}\\\textbf{Validation by\ldots}} \\
\midrule
\multirow{4}{*}{\rotatebox[origin=c]{90}{\emph{Academic}}}
  & \multirow{4}{*}{\makecell[l]{Canvas\\``course-\\ware''}}
  & \multirow{2}{*}{CS-Math courseware}             & 2  & SESMag    & Researchers
        & \multirow{9}{*}{\rotatebox[origin=c]{90}{Apply-<x>Mag}}
    & CS-Math instructor \\
  &                                           & & 2 & GenderMag & Researchers & & CS-Math instructor \\
\cmidrule(lr){3-6}\cmidrule(l){8-8}
  & & \multirow{2}{*}{CS-Arch courseware}           & 3 & SESMag    & Researchers & & CS-Arch instructor \\
  & &                                         & 3 & GenderMag & Researchers & & CS-Arch instructor \\
\cmidrule(lr){1-6}\cmidrule(l){8-8}
\multirow{5}{*}{\rotatebox[origin=c]{90}{\emph{Industry}}}
  & \multirow{5}{*}{\makecell[l]{Industry\\products}}
  & Digital-ID (RP)                           & 7
        & GenderMag & Digital-ID-RP team & & Team ID-RP member \\
  & & Digital-ID (ES)                         & 5  & GenderMag & Digital-ID-ES team & & Team ID-ES member \\
  & & Digital wallet                          & 3  & SESMag    & Wallet team        & & Entire wallet team \\
  & & Learning platform                       & 2  & SESMag    & Learning team      & & Entire learning team \\
  & & Municipal services                      & 6  & SESMag    & Municipal team     & & Entire municipal team \\
\bottomrule
\end{tabular}
}

\Description{Table describing evaluation contexts, products, and methods across Academic and Industry settings, all using Apply-‹x›Mag for Tool Evaluation.
Academic context, Type: Canvas "courseware": CS-Math courseware: 2 screens, Method SESMag, Manual Eval. by Researchers, Participant Validation by CS-Math instructor; CS-Math courseware: 2 screens, Method GenderMag, Manual Eval. by Researchers, Participant Validation by CS-Math instructor; CS-Arch courseware: 3 screens, Method SESMag, Manual Eval. by Researchers, Participant Validation by CS-Arch instructor; CS-Arch courseware: 3 screens, Method GenderMag, Manual Eval. by Researchers, Participant Validation by CS-Arch instructor.\\Industry context, Type: Industry products: Digital-ID (RP): 7 screens, Method GenderMag, Manual Eval. by Digital-ID-RP team, Participant Validation by Team ID-RP member; Digital-ID (ES): 5 screens, Method GenderMag, Manual Eval. by Digital-ID-ES team, Participant Validation by Team ID-ES member; Digital wallet: 3 screens, Method SESMag, Manual Eval. by Wallet team, Participant Validation by Entire wallet team; Learning platform: 2 screens, Method SESMag, Manual Eval. by Learning team, Participant Validation by Entire learning team; Municipal services: 6 screens, Method SESMag, Manual Eval. by Municipal team, Participant Validation by Entire municipal team.}
\end{table*}

\section{Empirical Evaluation\draftStatus{MMB 8/28: top d2.5}}
\label{sec:eval-design}

\boldification{Does it work? lets find out.}

\begin{figure*}[h]
    \includegraphics[width=0.9\linewidth]{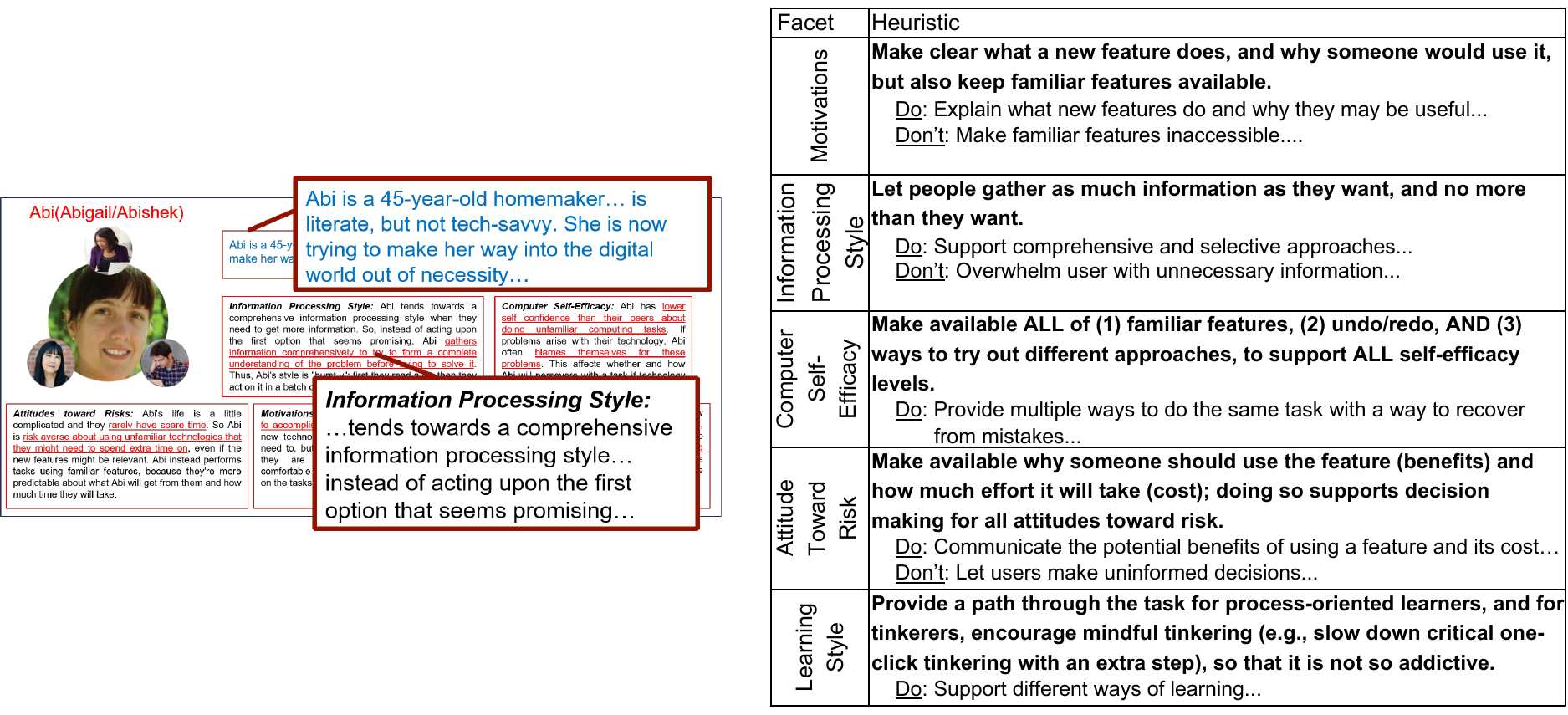}
    \caption{
    GenderMag overview.  GenderMag's facet endpoints are represented by Abi (left, portions magnified for readability) and Tim (not shown).  
    (Right:) The GenderMag facets and heuristics (unabbreviated in Supp. Docs).}
    \label{fig:gendermag}
    \Description{The GenderMag persona Abi, titled “Abi(Abigail/Abishek). Pictures of people of different genders and ethnicities are shown to represent Abi. Next to the picture is a description of Abi’s customized background, and below are Abi’s facet values. Callouts point to the description and to Abi’s Information Processing Style facet value. The description callout reads: “Abi is a 45-year-old homemaker… is literate, but not tech-savvy. She is now trying to make her way into the digital world out of necessity…” The Information Processing Style callout reads: “...tends towards a comprehensive information processing style… instead of acting upon the first option that seems promising…”}
\end{figure*}

How effective was Apply-<x>Mag?  
To find out, we compared manual <x>Mag bug-finding results by academic researchers and by industry product teams against the tool's automated bug-finding results.
As Table~\ref{tab:studyoverview} shows, the <x>Mags we used in our investigation were SESMag~\cite{burnett2024SESMag,busteed2026, chikezie2025SESmagSurvey} and GenderMag~\cite{burnett2016gendermag}.
We chose these <x>Mags to compare human (manual) results with the tool's because human precision with these is known to have
>=95\% precision~\cite{burnett2016gendermag, busteed2026,padala2020newcomersOSS, vorvoreanu2019genderMag}.
\footnote{Here, precision means \citet{mahatody2010walkthrough}'s measure: if a (manual) <x>Mag session identified an inclusivity bug in technology T, some user(s) later using technology T actually experienced that bug.
} 
SESMag's personas, facets, and heuristics were already discussed in Section~\ref{sec:tool}, and Figure~\ref{fig:gendermag} summarizes the GenderMag personas, facets, and heuristics.

Our empirical questions were:
\begin{itemize}
\item Q-UseCases: How effective is the tool at finding inclusivity bugs compared to human evaluators... (a)~in Use-case~1; (b)~in Use-case~2; (c)~in Use-case~4?  
\item Q-Context: In the two contexts we considered---an academic context in the U.S. and an industry context in India---did context matter to the tool's effectiveness?
\item Q-Environment: What are the environmental costs of using the tool, versus a practitioner chatting directly with an LLM evaluating the same screens?
\end{itemize}

The comparisons with manual results in two contexts (Table~\ref{tab:studyoverview}) served to evaluate Q-UseCases and Q-Context for Use-case~1 and Use-case~4, which both use <x>Mags already known to the tool.
In addition, we evaluated Use-case~2 computationally, by comparing results of Apply-<x>Mag as above, but without the tool already knowing these <x>Mags, i.e., the zero-shot version of Apply-<x>Mag. 
Use-case~3 did not require separate evaluation, as it is computationally the same as combinations of the other cases.
Finally, we investigated Q-Environment computationally.


\begin{figure*} [t]
    \centering
    \includegraphics[width=0.9\linewidth]{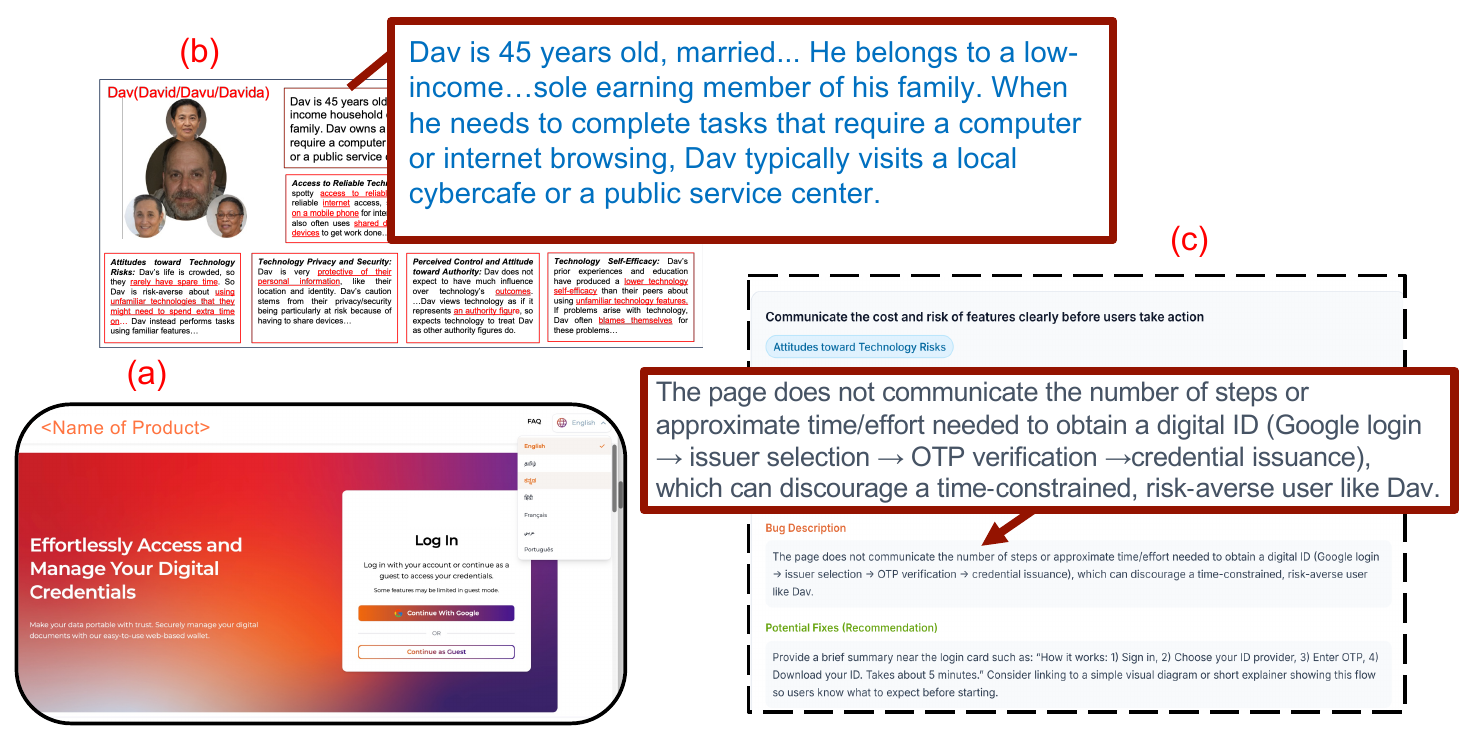}
    \caption{
    Team Wallet evaluated their (a) product using their (b) customized persona and then ran the same product through the tool, which generated (c) inclusivity bug reports like this one, in the same Why/Where/Fix format as in Figure~\ref{fig:bug_report}.} 
    \label{fig:teamwalletSESMAg}
    \Description{Three figures. Figure (a) is a screenshot of the landing page of a digital wallet website. Figure (b) shows the SESMag persona Dav, with a callout that reads: Dav is a 45 years old, married... He belongs to a low-income... sole earning member of his family. When he needs to complete tasks that require a computer or internet browsing, Dav typically visits a local cybercafe or a public service center.  Figure (c) is a screenshot of the tool output, with a callout pointing to the Bug Description that reads: The page does not communicate the number of steps or approximate time/effort needed to obtain a digital ID (Google login → issuer selection → OTP verification → credential issuance), which can discourage a time-constrained, risk-averse user like Dav.}
\end{figure*}

\subsection{Methodology\draftStatus{MMB 8/28/26: top is D3 (because it's empty) }}
\label{subsec:methods}

\subsubsection{Canvas "courseware" academic context\draftStatus{MMB 8/28/26 D2.4}}
\label{subsec:context1-method}

\boldification{We evaluated the tool on SESMag and GenderMag. The courses were...}
To answer our evaluation questions, we collaborated with academic instructors to apply the tool to Canvas ``courseware''. 
Canvas courseware was a particularly apropos choice because that type of technology has been used in prior inclusivity tool evaluations~\cite{chatterjee2022icer, chatterjee2024debugging}.
Two instructors offered their courses' courseware---CS-mathematics course (CS-Math), and CS-architecture course (CS-Arch).
The courses aimed at undergraduate Computer Science (CS) students enrolled remotely in U.S.-based courses.
Because students from lower-SES backgrounds and those historically underrepresented in computing struggle disproportionately in prerequisite (gateway) courses like these~\cite{cerf2022cs, hermann2022trinity}, they are good targets for inclusivity bug seeking.

\boldification{We needed ground truth, which was  researchers' manual evaluation, and checked the tool findings with the correspondign instructors}
For ground truth, we used human-identified inclusivity bugs from manual SESMag and GenderMag evaluations of this courseware.
Three academic researchers performed these sessions, including both author(s) and non-author(s) of this paper.
Working in pairs (two researchers/course), they manually used SESMag to find inclusivity bugs on five screens from the two CS courses.
The researchers then worked in different pairings to use GenderMag to evaluate those screens.
As is usual when using GenderMag or SESMag manually, they walked together with a persona through each scenario action-by-action, recording their answers to these two questions for every subgoal/action the persona ``should take'' (sample form in Supplemental Documents):
\begin{itemize}
    \item \textit{Before taking actions/subgoals}: Will <persona> have this subgoal/take this action? Why/what facets?
	\item \textit{After taking the ``should take'' action}: If <persona> does the right thing, will they know that they did the right thing and are making progress towards their goal? Why/what facets?
    \end{itemize}
As per prior literature, if the humans' answered ``maybe'' or ``no'' to any of these questions and \textit{ tied it to a particular <x>Mag facet value}, we counted it as an inclusivity bug, because that bug would \textit{disproportionately} affect individuals with that facet value~\cite{guizani2022whywherefix, burnett2016field, burnett2024SESMag, busteed2026}.

We then ran the tool on the same screens.%
\footnote{Note that, in contrast to the humans' manual process, the tool does not do <x>Mag walk-throughs; instead, it does <x>Mag heuristic evaluations.}
Finally, we conducted participant validation of all human- and tool-identified inclusivity bug reports through individual interviews with each course instructor, who agreed or disagreed with each report without knowing its source.
Each interview lasted 40--45 minutes. 

All human-participant evaluations followed IRB-approved protocols with informed consent. Participants received no monetary compensation.

\subsubsection{Industry context\draftStatus{MMB 8/28/26 D2.4 }}
\label{subsec:study2-method}

\boldification{The industry context differed from the academic context in ways that add generality to our investigation.}

The industry context added generality to our evaluation in several ways.
In this context, industry practitioners evaluated their own industry products using SESMag and GenderMag, which we then compared with the tool's SESMag and GenderMag bug-finding. 
Another difference from the academic context was that the products in the industry context were developed in India by Indian practitioners.
The industry products' target audiences ranged from teachers to Global South users of their products, whereas the academic courses' target audience was students enrolled in a U.S. university.

\boldification{we conducted an SESMag workshop and there we got the data for practitioners evaluating their products using SESMag walkthrough and our tool.}
The procedures were almost the same as in Section~\ref{subsec:context1-method}, but with the products' teams doing both the manual evaluations and the participant validations.
For the product teams' manual SESMag sessions, Company X invited industry practitioners to a hands-on workshop on using SESMag to evaluate their own products.
Four teams participated initially, with one dropping out early, leaving: the Wallet team, the Learning team, and the Municipal team (Table~\ref{tab:studyoverview}).
After training, each team independently evaluated its product for several hours using SESMag and a customized Dav persona appropriate for its product.
Figure \ref{fig:teamwalletSESMAg} shows examples from the Wallet team.
During the lunch break, we then ran the tool on their product screens.
When the teams returned, they reviewed the tool-identified inclusivity bug reports for their products and indicated which they considered valid.


The manual GenderMag data came from GenderMag evaluations Company~X software practitioners had done earlier to improve two of their products. 
For participant validation, we interviewed a member of each product team following the same process as in Section~\ref{subsec:context1-method}. 

\subsubsection{Qualitative coding for bug-report matching\draftStatus{MMB 8/31 D2.2 but a bit long}}

\boldification{To start with comparing the tool's outputs with the manual evaluation data, why comments from manual evaluations have been labeled with bug categories and got IRR>=80\% }

To compare the tool's bug-finding with the humans' manual bug-finding, we needed a way to match the tool's and humans' inclusivity bug reports.
When a configuration includes a bug-category list, the tool uses the list to inform its search and labels each report with the matched bug category/ies (Section~\ref{sec:tool}).
The SESMag and GenderMag configurations included such lists, so the tool had already ``coded'' all its reports (e.g., lack of guidance, unexplained jargon; complete codebook in Supp. Docs.).

We used the same bug category lists to code the humans' manual reports.
Two researchers independently coded 20\% of the humans' inclusivity bug reports for SESMag and GenderMag, achieving inter-rater agreement (IRR) of 82\% and 81.5\%, respectively, using the Jaccard index~\cite{landis1977measurement}.  
One researcher then finished coding the remaining inclusivity bug reports. 
This enabled us to align the inclusivity bug reports for each screenshot by matching the tool-reported bug categories with those coded for the manual findings on that screen.

\subsubsection{Other use-case evaluations\draftStatus{D2 Amreeta: 9/8/26}} 

\boldification{We compared tool output with zero-shot version of the tool}
We then evaluated Use-case~2 computationally, which concerns applying the tool to a new <x>Mag. Although ADHD-Mag motivated this use case, ADHD-Mag does not yet exist. We therefore evaluated Use-case~2 using SESMag: we ran the tool on the same academic courseware materials used in Use-Case~1, but in zero-shot mode~\cite{syed2024zeroshot}, without providing worked examples. We compared the results with the Use-case~1 SESMag run. This evaluated Use-case~2's viability while isolating the importance of worked examples for the tool's effectiveness. 


\boldification{We evaluated intersectional bug-finding}
Lastly, we evaluated Use-case~4, which supports intersectional bug-finding.
As explained in Section~\ref{subsec:Use-case4}, Fallatah et al.~\cite{fallatah2025intersectionalmag} showed that with analytical methods like those in the InclusiveMag family, the union of bugs found by practitioners using different <x>Mags is at least as accurate as the bugs found by practitioners directly analyzing an intersectional population. 
Using scenarios  with both SESMag and GenderMag results, we leveraged this insight as follows:
We compare Apply-<x>Mag's Use-case~1 SESMag $\cup$ GenderMag bugs found against running Apply-<x>Mag compositionally on SESMag$\times$GenderMag, as per the Use-case~4 compositional approach. As already mentioned, we did not directly evaluate Use-case~3, since it combines the operations evaluated in Use-Cases~1 and~2. 

\subsubsection{Why not just chat with an LLM? An environmental evaluation\draftStatus{D2 (Amreeta 9/8 5:42pm)}}
\label{subsubsec:method-environment}

\boldification{To answer Q-Environment, we measured the tool's prompting (Box~3b + Box~5) environmental cost. compared versus a practitioner chatting directly with an LLM to evaluate the CS-Math courseware---but GreenQuery only sees a browser, so we ran both in the browser.}
%
To answer Q-Environment, we used the GreenQuery Chrome extension~\cite{ngai2026greenquery} to compare the environmental costs (water and CO$_2$) of using Apply-<x>Mag with a practitioner chatting directly via ChatGPT.
We entered the practitioner chatting prompt directly into ChatGPT GPT-5.6 Sol
\footnote{We had to use GPT-5.6 Sol because GPT-5.1 (the tool's current LLM) was not available on the ChatGPT website~\cite{openai2026chatgptreleasenotes}.
}. 
Since GreenQuery requires browser interaction, we simulated Apply-<x>Mag using GPT-5.6~Sol by harvesting Apply-<x>Mag's prompts from our logs and entering them directly in the browser (see the Supp. Docs for sample tool prompts).

\boldification{Here is what each Q-Env evaluation actually got---the chatting prompt was not just a ``make it inclusive'' request, the LLM recieved the same materials the tool works from.}
The ChatGPT prompt was as a practitioner might enter if they were well informed about prompting LLMs about <x>Mags, without iterative clarifications. 
That prompt included the same information as the tool prompt---the heuristics, the persona's facet values, customized persona background, the scenario, and the screen---all in one prompt. For example, the human's chat prompt to run SESMag on a screen was:

\borderedbox{
Detect SES inclusivity bugs on the image I uploaded. use the csv file containing the SESMag heuristics and DAV persona facet values\ldots
<cut for brevity: 7 more paragraphs enumerating the Dav persona facets, customized background, scenario description, plus the screen image and spreadsheet as attachments; full prompt in Supp. Docs.>}

We ran this comparison five times on the two CS-Math courseware screens with SESMag and then with GenderMag, and averaged the results.
After each run, we cleared the LLM's history so that no measurement depended (much) on any automatic caching by the LLM.  
All prompts used the zero-shot <x>Mag configurations (i.e., with no worked examples).

%


\subsection{Use-case~1 results for using Apply-<x>Mag on existing <x>Mags\draftStatus{top D2: Amreeta 9/6 1 PM}}
\label{sec:rq1}
\label{subsec:results-Use-case1}

\boldification{Use-case 1 Story: The tool agreed with humans a lot and behaved similarly for both Mags.}

Recall that in Use-case 1, the practitioner invokes the tool on <x>Mags it has seen before.
We investigated this use-case in both the academic context and the industry context.

In the academic context, when Apply-<x>Mag applied SESMag and GenderMag to the courseware, it produced 28 inclusivity bug reports, which matched the humans 25 times, as detailed in the top half of Table~\ref{tab:TP+FP+FN-totals}.
Apply-<x>Mag's inclusivity bug reports usually matched the humans' judgments (25/28=89\% true positives), but humans did not agree with the remaining 3 inclusivity bug reports (3/28=11\% false positives).
\footnote{Here, we treat the humans' decisions as the ``ground truth'', as per the humans' manual inclusivity bug reports and/or the instructors' validations, because research has shown that humans' GenderMag and SESMag results to have >=95\% precision rates~\cite{burnett2016gendermag, busteed2026,padala2020newcomersOSS, vorvoreanu2019genderMag}.} 
The tool also missed 3 bugs the human found (3/28=11\% false negatives).

\begin{table}[h]
\centering
\caption{
Inclusivity bug report results by screen (courseware) and by industry product (with screen-by-screen details in Supp. Doc.) 
The 119 bug reports are the union of 115 human-generated bug reports and 101 tool-generated bug reports. In total, Apply-<x>Mag found 97/115 (84\%) of the bugs humans found, missed 18/115 (16\%), and incorrectly flagged 4.
where: \\
\emph{TP=True Positives} = tool matched humans;
\emph{FN=False Negatives} = tool missed bugs the humans found;
\emph{FP=False Positives} = tool found but humans disagreed.
}
\label{tab:TP+FP+FN-totals}
\resizebox{\linewidth}{!}{%
\setlength{\tabcolsep}{4pt}
\begin{tabular}{@{} l l *{6}{c}ccc @{}}
\toprule [1pt]
 & & \multicolumn{3}{c}{\textbf{SESMag}} & \multicolumn{3}{c}{\textbf{GenderMag}} & \multicolumn{3}{c}{\textbf{Totals}} \\
\cmidrule(lr){3-5} \cmidrule(lr){6-8} \cmidrule(lr){9-11}
\textbf{Course} & \textbf{Screen} 
  & TP & FP & FN
  & TP & FP & FN
  & TP & FP & FN \\
\midrule 
\multirow[c]{2}{*}{{\footnotesize\textbf{CS-Math~}}} 
  & Homepage          & 3 & 0 & 0 & 2 & 0 & 0 \\
  & Modules         & 3 & 0 & 0 & 3 & 0 & 2 \\
\cmidrule{1-8}
\multirow[c]{3}{*}{{\footnotesize\textbf{CS-Arch}}}
  & Homepage         & 3 & 0 & 0 & 1 & 1 & 0 \\
  & AssignDescrip\     & 2 & 1 & 0 & 3 & 0 & 0 \\
  & AssignSubmit\    & 3 & 1 & 1 & 2 & 0 & 0 \\
\midrule
\textbf{Courseware total}
  & {31 bug reports} & \textbf{14} & \textbf{2} & \textbf{1} & \textbf{11} & \textbf{1} & \textbf{2} 
  & \textbf{25} & \textbf{3} & \textbf{3 }\\
\midrule[1pt]
  \footnotesize\textbf{Digital wallet}       & 3 Screens & 13 & 0 & 2 & -- & -- & -- \\
  \footnotesize\textbf{Learning platform}    & 2 Screens & 6  & 1 & 1 & -- & -- & -- \\
  \footnotesize\textbf{Municipal services}   & 6 Screens & 23 & 0 & 2 & -- & -- & -- \\
\cmidrule{1-8}
 \footnotesize\textbf{Digital ID-RP}     & 7 Screens   & -- & -- & -- & 16 & 0 & 7 \\
 \footnotesize\textbf{Digital ID-ES}     &  5 Screens  & -- & -- & -- & 14 & 0 & 3 \\
\midrule
\textbf{Industry total}
  & {88 bug reports} & \textbf{42} & \textbf{1} & \textbf{5} & \textbf{30} & \textbf{0} & \textbf{10} 
  & \textbf{72} & \textbf{1} & \textbf{15} \\
\midrule[1pt]
\textbf{Grand total}
  & {119 bug reports} & \textbf{56} & \textbf{3} & \textbf{6} & \textbf{41} & \textbf{1} & \textbf{12} 
  & \textbf{97} & \textbf{4} & \textbf{18} \\
\bottomrule [1pt]
\end{tabular}
}

\end{table}

The industry context results were similar but not exactly the same.
In the industry context, the tool's true positives rates and false positives rates were a little better, but its false negative rate was a little worse.
As the bottom half of Table~\ref{tab:TP+FP+FN-totals} shows, the tool matched the human teams all but once---only 1/73=1\% false positives---but missed 15/87 of the bugs the humans found (17\% false negative).
Thus, the answer to Q-Context is that context made a small difference.

\begin{table}[h]
\caption{Precision, recall, and F1 scores by <x>Mag, for the courseware and real-product datasets. \textit{Precision} = TP/(TP+FP); \textit{Recall} = TP/(TP+FN); \textit{F1-score} = 2(Precision$\times{}$Recall)/(Precision+Recall). The ``Overall (weighted average)'' value is weighted by the number of screens processed.}
\label{tab:qMetricsCombined}
\centering
\setlength{\tabcolsep}{4pt}
\resizebox{\linewidth}{!}{%
\begin{tabular}{@{}llccc@{}}
\toprule
 & & \textbf{SESMag} & \textbf{GenderMag} 
 & \textbf{\shortstack{Weighted \\average}} \\
\midrule
\multirow{3}{*}{\textbf{Courseware}}
 & \textbf{Precision}          & 88\% & 92\%  & 90\% \\
 & \textbf{Recall}             & 93\% & 85\%  & 89\% \\
 & \textbf{F1-score (as \%)}   & 90\% & 88\%  & 89\% \\
\midrule
\multirow{3}{*}{\textbf{Industry}}
 & \textbf{Precision}          & 98\% & 100\% & 99\% \\
 & \textbf{Recall}             & 89\% & 75\%  & 82\% \\
 & \textbf{F1-score (as \%)}   & 93\% & 86\%  & 89\% \\
\bottomrule
\end{tabular}
}
\end{table} 

\boldification{both precision and recall compare favorably with prior AID stats (use both AIDs), not quite as good as Jon on precision but way better on recall.}

These rates produce the precision, recall and F1 results shown in Table~\ref{tab:qMetricsCombined}.
Apply-<x>Mag's overall precision (90\%--99\%) and recall (82\%--89\%) compare very favorably with previous tools that automated GenderMag (the only <x>Mag with prior automation attempts). 
AID~\cite{chatterjee2021aid} automated a subset of GenderMag for open source project sites, which produced only 69\% precision but an impressive 92\% recall, for an overall F-measure of 79\%.
A later version of AID that evaluated online courseware for the full GenderMag~\cite{chatterjee2024debugging} reported an 82\% IRR among researchers. 
Finally, when \citet{geuenich2026aicanseewhatyoucant} used an LLM agent to automate GenderMag evaluations of several web interfaces, they reported precision ranging from 83\%--100\%, but low recall of <=35\%.
Note also that Apply-<x>Mag's precision and recall performance worked for both SESMag and GenderMag with similar results, whereas the AID and Geuenich et al. systems were crafted especially for GenderMag.

\boldification{we care the most about precision because... Still, both precision and recall is awesome.}
Of these two metrics, precision is the higher priority for us.
Informally, precision is about not lying---so Apply-<x>Mag's high precision suggests that  practitioners can believe much of what the tool reports.
Fortunately, Apply-<x>Mag also did reasonably well on recall.
An interpretation of its 82\%--89\% recall is that practitioners can regard the tool's SESMag and GenderMag evaluations on products like these as being reasonably complete.
That is, even if they supplemented the tool's reports with humans doing a manual, they would probably not find many more inclusivity bugs.


\subsubsection{Examples of hits and misfires\draftStatus{MMB 9/5/26 d2ish}}
\label{subsubsec:hitsAndMisfires}

\boldification{When the tool's bug finding matched human judgement, we'll call it a hit.  For example...}

When the tool found a bug that human(s) agreed with (the 97 true positives), we can say it ``hit'' the target. 
One such hit was on the CS-Math homepage shown earlier in Figure~\ref{fig:sample-input-screenshot}.
As its Why/Where/Fix inclusivity bug report explained (likewise shown earlier in Figure~\ref{fig:bug_report}), the tool considered that inclusivity bug to disproportionately affect low-self-efficacy users like SESMag's Dav, and suggests several ways the page could add scaffolding to fix that inclusivity bug. 
Another hit was on Team Wallet's ``pick a language'' page, seen earlier in Figure~\ref{fig:teamwalletSESMAg}.
The tool considered the inclusivity bug to disproportionately affect risk-averse users like SESMag's Dav. 
The Why/Where/Fix bug report focused on the particular risk of wasting extensive time, which users like Dav cannot afford.
Its suggested fix was to clarify exactly how it works and how long it should take.

\begin{figure*}[h]
    \includegraphics[width=0.9\linewidth]{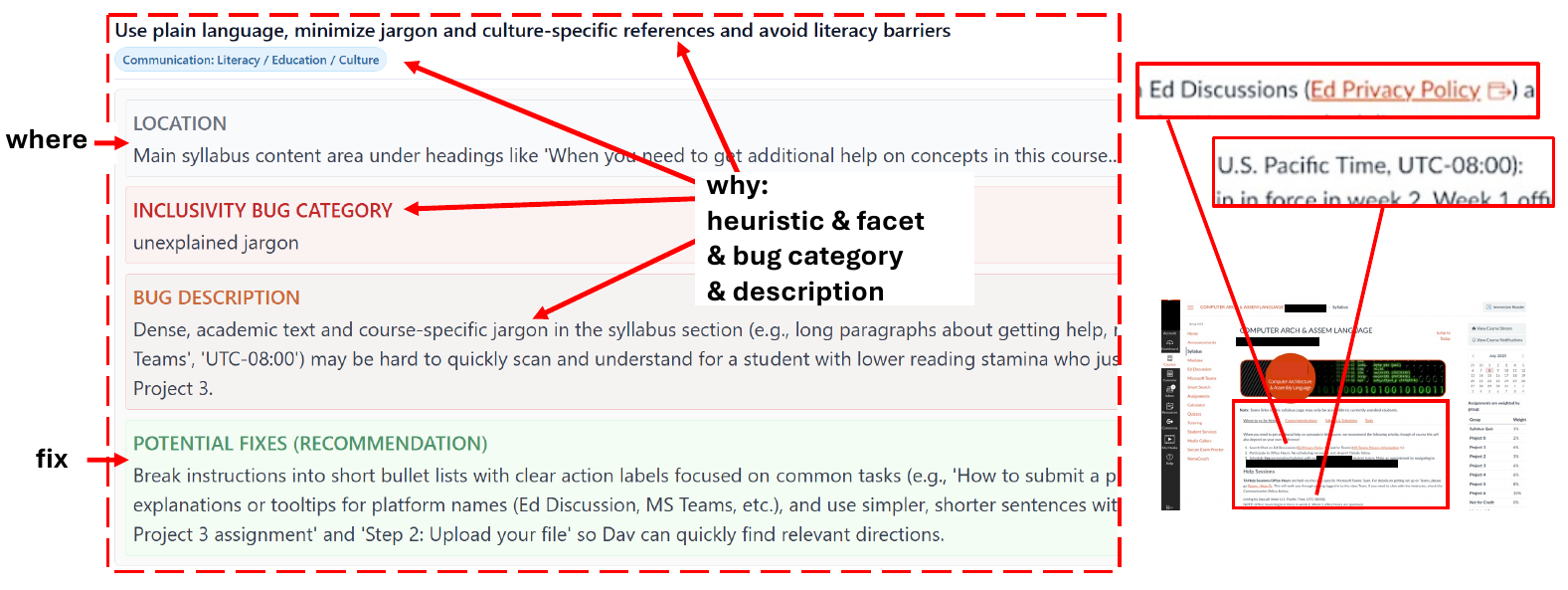}
    \caption{Inclusivity bug report by Apply-<x>Mag that a CS-Arch instructor validated on the course homepage with callouts. Here, the bug occurred in the main syllabus content (bottom right \textcolor{red}{red} box).}
    \label{fig:bugreport-unexplainedjargon-csArch}
    \Description{A screenshot of the tool-generated inclusivity bug report. The title reads: Use plain language, minimize jargon and culture specific references and avoid literacy barriers. Just below the title is a blue tag labeled with the facet: Communication: Literacy / Education / Culture. Below are the location, inclusivity bug category, bug description, and potential fixes. LOCATION: Main syllabus content area under headings like ‘When you need to get…” <cut off>. INCLUSIVITY BUG CATEGORY: unexplained jargon. BUG DESCRIPTION: Dense, academic text and course-specific jargon in the syllabus section (e.g., long paragraphs… <cut off> …quickly scan and understand for a student with lower reading stamina who just wants to know… <cut off>. POTENTIAL FIXES (RECOMMENDATION): Break instructions into short bullet lists with clear action labels focused on common tasks (e.g.,...) <cut off> …(Ed Discussion, MS Team, etc.), and use simpler shorter sentences with headings like ‘Step-1:... <cut off> … directions.\\In the top right corner, an annotation with arrows show how the report follows Guizani et al.’s Why/Where/Fix approach. “Why” points to the title. “Where” points to the LOCATION box. “Heuristic & Facet” points to the facet tag under the title. “Bug category” points to the INCLUSIVITY BUG CATEGORY box. “Description” points to the BUG DESCRIPTION box. “Fix” points to POTENTIAL FIXES (RECOMMENDATION) box.}
\end{figure*}

\boldification{More interesting hits were when tool found bugs the researchers had missed.}
The most interesting hits were when Apply-<x>Mag found an inclusivity bug that the evaluators' initial evaluations had missed, but then the after-the-tool-run participant validators affirmed.
In the academic context, this happened only once, when Apply-<x>Mag's application of SESMag  spotted some academic jargon on a CS-Arch class page (Figure~\ref{fig:bugreport-unexplainedjargon-csArch}).
The CS-Arch instructor found this inclusivity bug to be very actionable, and started brainstorming fixes right away, ``\textit{I could probably add context to that... highlight ED Discussions in the course discussion board.}''
In the industry context, this type of hit was very common, arising 31 times.
One example arose with the Digital ID-RP team's \textit{Manage virtual ID} screen, where the tool reported unclear action requirements on the download and delete icons---which the human team's GenderMag session had originally missed, but was then verified by a Digital ID-RP team member when they saw the inclusivity bug report: ``\textit{...don't look clickable. <Abi> might not click.}''
These kinds of hits are consistent with \citet{geuenich2026aicanseewhatyoucant}'s results: sometimes ``AI can see what you can’t see''.  

\begin{table*}[h]
\centering
\caption{
Tool hits and misfires. TP, FP, FN, as per Table~\ref{tab:TP+FP+FN-totals}. The rightmost TP column (Tool first...) shows the ``AI can see what you can't'' hits (32 occurrences).  The rightmost (FN) column (Tool missed...) shows the opposite: humans seeing what AI can't (18 occurrences).
}
\label{tab:hits-misses}
\resizebox{0.8\linewidth}{!}{%

\setlength{\tabcolsep}{2pt}
\begin{tabular}{@{}lccccc@{}}
\toprule [1pt] 
 {} & \multicolumn{2}{c}{\textbf{Hits (TP)}} & {}  & \multicolumn{2}{c}{\textbf{Misfires (FP+FN)}}  \\
\cmidrule(lr){2-3} \cmidrule(lr){5-6}
\textbf{}
      & \makecell[c]{\textbf{Humans}\\\textbf{= tool}}
      & \makecell[c]{\textbf{Tool first,}\\\textbf{humans agreed}}
      & {}
      & \makecell[c]{\textbf{Tool only, humans}\\\textbf{disagreed (FP)}}
      & \makecell[c]{\textbf{Tool missed}\\\textbf{humans found (FN)}} \\
\midrule
\textbf{Academic context}         & 24 & 1  & {} & 3  & 3  \\
\addlinespace
\textbf{Industry context}         & 41 & 31 & {} & 1  & 15 \\
\midrule
\midrule
\textbf{Totals}
  & 65 & 32 & {} & 4 & 18 \\
\midrule [1pt] 
\textbf{Tool's hits and misfires}
  & \multicolumn{2}{c}{\textbf{97 hits}} & {} & \multicolumn{2}{c}{\textbf{22 misfires}} \\
\bottomrule [1pt] 
\end{tabular}
}

\end{table*}
\begin{table}[ht]
\caption{
The same tool, prompted zero-shot instead of few-shot, on the academic context courseware screens. \textit{Zero-shot (Use-case~2)} = how many of the tool's Use-case~1 bug reports (Table~\ref{tab:TP+FP+FN-totals}) the Use-case~2 run also produced. Removing the worked examples lost two SESMag bug reports but none of GenderMag's.}
\label{tab:zeroshot-comparison}
\centering
\begin{tabular}{lrrr}
\toprule
 & \textbf{Use-case~2} & \textbf{Use-case~1} & \textbf{Agreement} \\
 & \textbf{(zero-shot)}  & \textbf{(few-shot)}  & \\
\midrule
SESMag           & 14 & 16 & 88\%  \\
GenderMag        & 12 & 12 & 100\%   \\
\midrule
\textbf{Overall} & 26 & 28 & 93\%  \\
\bottomrule
\end{tabular}
\end{table}


\boldification{turning to misfires...}

The opposite was also sometimes true: sometimes humans could understand things Apply-<x>Mag could not.
Most of the 22 tool misfires (4 false positives + 18 false negatives) fell into three categories: (1)~outright buggy behavior by the tool, (2)~missing understanding of human expectations about sequences of screens, or (3)~professional judgment that the humans had and the tool did not.

The first category was at the surface level, with the tool misreading something on the screen.
For example, this arose twice when the tool complained that paragraphs should be bullet lists, when in fact they were already bullet lists.
It also arose once when the tool thought that two colors were identical, which confused it into erroneously reporting that a link did not seem clickable.

The second category went deeper.
In the second category, the tool would have needed to reason about user expectations that arise from a \textit{sequence} of screens and actions.
For example, the humans using GenderMag to manually evaluate the site found an inclusivity bug tied to Abi's expectations that a navigation from ``Start here'' would lead to information about Week~1: ``Abi sees nothing related to Week 1 so they feel little progress.''
As information foraging theory research has shown, when users follow the ``scent'' of potentially useful information, they expect to move from general information to specific~\cite{spool1998infoscent}.
The human evaluators explained the inclusivity bug in a way consistent with this: users like Abi learn new technologies best using organized processes, and here their expectations would not be met.
Reasoning about such scent-following sequences is outside the tool's abilities.
Instead, it does its specialized heuristic evaluation one screen at a time. 
Although it might gain some information from what it has seen on previous screens, it does not attempt to reason explicitly about sequences.

The third category was about professional judgment that humans build over time.
The clearest example arose in two CS-Arch inclusivity bug reports, when Apply-<x>Mag said that course assignment descriptions were too dense, long, and/or complex: 
\borderedbox{
The long assignment description and requirements are presented as a dense wall of text with many numbered and bulleted lists and no clear visual separation between what is essential to submit Project 3 (e.g., how/where to submit) and what is background or optional reading. This makes it hard for ...
}

However, the CS-Arch instructor had taught this course for quite awhile, and had learned from experience:
``\textit{I don't <agree>... I used to have way less information on these assignment pages, and people had ...overwhelmingly more trouble.}''

\boldification{we corroborate jon's paper, but also don't forget our awesome precision \& recall}

Together, the hits and misfires show that our results corroborate what \citet{geuenich2026aicanseewhatyoucant} reported: First, that AI can find what human evaluators miss, and conversely, human evaluators can find what AI misses (Table~\ref{tab:hits-misses}). 
That said, Apply-<x>Mag filled in many more gaps than the humans did (32 vs. 18), but running both gave the most complete set of results.
We further discuss pluses and minuses of the tool vs. manual in the Discussion section.


\subsection{Use-case 2: Was zero-shot-Apply-<x>Mag reasonably good?\draftStatus{D2.2 (MMB 9/3/2026 11 am)}}
\label{subsubsec:zeroshot}
\label{subsec:results-Use-case2}


\boldification{Even without worked examples, the tool still found 26 of the same 28 inclusivity bug reports---which matters because a practitioner adding a new <x>Mag doesn't have worked examples to give it.}
Recall that Use-case~2 did not have the benefit of worked examples, which meant the tool had to switch to a zero-shot prompt instead of few-shot. 
Fortunately, without worked examples, the tool was still reasonably effective.
With the zero-shot prompt, it still found 26 of the same 28 inclusivity bug reports it found previously on the courseware (Table~\ref{tab:zeroshot-comparison}).
The agreement between the Use-case~2 (zero-shot) and Use-case~1 (few-shot) tool runs (93\%) was similar under both <x>Mags---the zero-shot run matched 14/16 inclusivity bug reports for SESMag and 12/12 for GenderMag.
This demonstrates the viability of Use-case 2 (Section~\ref{subsec:Use-case2}), where a practitioner uses the tool to enter and then run a new <x>Mag they did not invent themselves.


\subsection{Use-case 4: Intersectional results\draftStatus{D1 (MMB 9/10, 2252: d2.25)}}
\label{subsec:results-Use-case4}

\boldification{Composing the two Mags into one intersectional Mag found the same bugs as running them separately and union'ing the results (serial)---same 25 found, same 3 missed.}
Recall from Section~\ref{subsec:Use-case4} that the serial and compositional approaches to intersectional <x>Mag should surface the same bugs, and that the validity of union'ing single dimension bugs found analytically  is established in prior work~\cite{fallatah2025intersectionalmag}.
Here we consider whether Apply-<x>Mag worked in exactly this way---i.e., whether combining SESMag and GenderMag results serially (one after the other, then taking the union), produced the same results as running SESMag$\times$GenderMag all at once.

\begin{table*}[!t]
\caption{Environmental costs of Apply-<x>Mag versus a human chatting with an LLM on two CS-Math courseware screens, averaged over five runs. \textbf{Bold} indicates the lower environmental cost in each row. Apply-<x>Mag was less resource-intensive in nearly every comparison.}
\label{tab:qenvironmentCost}
\resizebox{0.8\linewidth}{!}{%
\begin{tabular}{llrrrr}
\toprule
&  & \multicolumn{2}{c}{\textbf{Apply-<x>Mag}} & \multicolumn{2}{c}{\textbf{Human-LLM chat}} \\
\cmidrule(lr){3-4} \cmidrule(lr){5-6}
\textbf{<x>Mag} & \textbf{Screen (prompt)} & Water\,(L) & CO$_2$\,(g) & Water\,(L) & CO$_2$\,(g) \\
\midrule
\multirow{5}{*}{\textbf{SESMag}}
 & One-time (send config, get detection-tactic) & 0.014          & 4.5               & \textbf{--}    & \textbf{--}  \\
 & Homepage (analysis)                          & \textbf{0.002} & \textbf{0.6}      & 0.008          & 2.7 \\
 & Modules (analysis)                           & \textbf{0.003} & \textbf{0.8}      & 0.008          & 2.4 \\
 \cmidrule(l){2-6}
 & \textbf{Subtotal:} 2 screens' analysis prompts  & \textbf{0.005} & \textbf{1.4}   & 0.016          & 5.1 \\
 & \textbf{Total:} all prompts                     & 0.019          & 6.0            & \textbf{0.016} & \textbf{5.1} \\
\midrule
\multirow{5}{*}{\textbf{GenderMag}}
 & One-time (send config, get detection-tactic)   & 0.011          & 3.6 & \textbf{--}   & \textbf{--}  \\
 & Homepage (analysis)                            & \textbf{0.002} & \textbf{0.6} & 0.008 & 2.4 \\
 & Modules (analysis)                             & \textbf{0.001} & \textbf{0.4} & 0.007 & 2.3 \\
 \cmidrule(l){2-6}
 & \textbf{Subtotal:} 2 screens' analysis prompts  & \textbf{0.003} & \textbf{1.0} & 0.015 & 4.7 \\
 & \textbf{Total:} all prompts                     & \textbf{0.014} & \textbf{4.6} & 0.015 & 4.7 \\
\bottomrule
\end{tabular}
}

\end{table*}


The results turned out to very similar, but not exactly the same.
In both approaches Apply-<x>Mag found the same 25 of the 28 human-reported bugs previously shown in Table~\ref{tab:hits-misses} (``Courseware total'' row, rightmost TP column), and also missed the same 3 (rightmost FN column). 
However, the compositional SESMag$\times$GenderMag found one more bug than in Table~\ref{tab:hits-misses}.

When we saw this one-bug difference, we suspected the cause to be LLMs' generative variability~\cite{ho2024application}.
To test this hypothesis, we ran the compositional SESMag$\times$GenderMag yet again---and received still another slightly different total.
As this difference shows, Apply-<x>Mag's dependence on an LLM means it cannot fully assure that it will produce exactly the same answers serially as  compositionally, or even exactly the same answers in two runs done exactly the same way.
Still its results consistently compared very well with humans' in every use-case, suggesting that even though slight variability will occur, the heuristics and/or examples and detection tactics give the LLM enough guidance to keep its answers fairly similar.

An advantage of the compositional approach over the serial approach lay in how the compositional approach sometimes brought both applicable SESMag facet values and applicable GenderMag facet values together to explain a bug. 
For example, in the compositional approach it attributed a bug in CS-Math's Modules page to a combination of SESMag's Communication facet with GenderMag's Motivations facet.
The bug report's second underlined phrase below (added for clarity) shows the connection to SESMag/Dav's lower tolerance for reading extensive passages (Communication facet), and the first underlined phrase (also added) shows the connection to GenderMag/Abi's Motivations facet value (Abi's task-oriented motivation to take action toward her goal).

\borderedbox{
The modules and items are presented as homogenous lines of text with small document icons, with \ul{no visual emphasis on Week 1 tasks}, deadlines, or priorities. Students must read each item rather than relying on visual cues, which \ul{increases cognitive load for those who benefit from visual structure}.}

\subsection{Environmental evaluation results\draftStatus{D2.4 (FAM 9/9 6pm)}}
\label{subsec:results-environment}


\fixme{FAM}{stick these somewhere as a reminder, the methods to this results is far readers may forget**
Water (L) and carbon emissions (g CO$_2$) are GreenQuery~\cite{ngai2026greenquery} estimates of a browser session on GPT-5.6 Sol, as described in Section~\ref{subsubsec:method-environment}.
}

\boldification{In Use-case 1 the tool has no start-up cost at all, so the only comparison is per screen, the human-LLM chat cost between 3 to 5 times as much as Apply-<x>Mag, for both <x>Mags.}
To compare Apply-<x>Mag's environmental cost to a human-LLM chat, we considered Use-cases~1 and~2 separately.
For Use-case~1, in which Apply-<x>Mag has already seen the <x>Mag(s), it did not need to generate detection tactics.
As Table~\ref{tab:qenvironmentCost}'s 2-screen subtotals show, Apply-<x>Mag's costs averaged only 0.003--.005\,L water and 1.0--1.4\,g CO$_2$ for GenderMag and SESMag, respectively.
In contrast, a human chatting directly with an LLM cost 3--5 times as much water and CO$_2$.


\boldification{Why is human-LLM chat more expensive? Because every chatting prompt has to carry the configuration again.}
The human-LLM chat's main expense, unlike Apply-<x>Mag's, was informing the LLM of the <x>Mag configuration with every chatting prompt.
This assumes separate human-LLM chat sessions for each screen---but even if the second (Modules) prompt approached ``free'' (because of LLM-cached information in the same session), the human-LLM chat would still cost more than Apply-<x>Mag's 2-screen subtotal (e.g., for GenderMag, 1.0\,g CO$_2$ for Apply-<x>Mag two screens vs. 2.4\,g CO$_2$ for human-LLM chat Homepage only, with Modules screen being ``free'').
\footnote{Even if the practitioner saved their LLM project context forever, Apply-<x>Mag would still save resources in the long run, because its global storage is available to \textit{all} future users of that <x>Mag.
}

\boldification{Adding a new <x>Mag costs one extra prompt up front, but it paid itself off within two or three screens---and after detection-tactic prompt is run once it is never run again (i.e., pay once).}
Use-case~2, in which Apply-<x>Mag has not seen the new <x>Mag before, imposes a front-loaded cost. 
When a practitioner provides a new <x>Mag configuration, the tool must get the detection-tactics from the LLM---but only once.
It stores the configuration and detection tactics globally, so it never needs that prompt again unless the configuration changes (e.g., Use-case~3).
Our results show that this front-loaded cost was quickly repaid.
As Table~\ref{tab:qenvironmentCost}'s Totals show, in SESMag's case, Apply-<x>Mag's Total would break even with the human-LLM chat after three screens, and in two screens GenderMag's total was already lower.
Any future analyses using SESMag or GenderMag uses only the analysis prompts (not the detection-tactic prompt), since these <x>Mags' configurations and detection tactics are now stored.

\boldification{And both of these costs are \textit{tiny} compared compared to everyday technology activities.}
Compared to everyday technology activities, Use-case~1's and Use-case~2's costs were small.
For example, one hour of Zoom costs 2--12\,L of water and 150--1,000\,g of CO$_2$~\cite{morris2023zoomCost}, >100 times more water and >25 times more CO$_2$ than Apply-<x>Mag applying SESMag (Table~\ref{tab:qenvironmentCost}'s total).
Or, in terms of Google-searches, Apply-<x>Mag's analysis prompt's environmental cost about the same as 2--4 non-AI Google searches (0.2\,g CO$_2$ each)~\cite{greenspector2020search, pawade2023digitalFootprint}.


\section{Discussion}
\label{sec:discussion}

\subsection{Reasons to use the tool and reasons not to\draftStatus{9/9/26 MMB d2.4, FAM says D3 9/10 11:40pm}}

To this point, we have presented Apply-<x>Mag from somewhat of a ``tooling up is good'' perspective.
In this section we weigh what practitioners and researchers stand to gain against what they stand to lose by using the tool.

The tool's advantages lie in saving time and, in some situations, environmental savings.
Regarding time, our results show that using Apply-<x>Mag can eliminate the labor-intensiveness of manually using <x>Mag, which \citet{chatterjee2022icer} reported to be about 2 hours  for 1-3 scenario evaluations.
Regarding environmental savings, it depends.
Compared to a practitioner chatting with an LLM directly, Section~\ref{subsec:results-environment} showed Apply-<x>Mag to be much less costly in environmental resources as soon as an <x>Mag becomes known to the tool.
Still, compared to a co-located team working together, manual evaluation would be less environmentally costly.
However, the opposite is true for a distributed team. 
Recall from Section~\ref{subsec:results-environment} that Apply-<x>Mag would use fewer environment resources than a distributed team doing a manual <x>Mag via zoom due to the environmental cost of a zoom session.

\boldification{**NOT-TO-USE reasons: social climate + defensiveness}

However, there are also many reasons not to use Apply-<x>Mag.
When humans do <x>Mag evaluations manually, prior literature has shown a plethora of sociotechnical benefits.
One such benefit is social climate. 
Research shows that development teams' manual use of an <x>Mag can lead to increased appreciation of each others' facet value differences (e.g., ``\textit{<My teammates understand that we> tend to work differently <and thus we were> less demanding on each other}''~\cite{letaw2021onlineCS}), and even  statistically significant improvements in their ratings of team climate~\cite{garcia2023regularCS}.  
Research also reports that doing <x>Mags manually can decrease developers' defensiveness about their design decisions~\cite{hilderbrand2020engineering, chattopadhyay2025MOSIP}.  
As one of \citet{hilderbrand2020engineering}'s participants put it: ``\textit{we have the Devs who designed this UI and it was like once they were <persona> they could let go of their ego.}''

\boldification{**more NOT-TO-USE reasons: empathy }

Social benefits of continuing to <x>Mag manually can extend to the ultimate end users.  
For example, 
``\textit{if I didn’t have access to reliable technology like Dav, I would want to ...}''~\cite{busteed2026};
``\textit{... putting myself in the shoes of others really helped me to understand the shortcomings of our software}''~\cite{garcia2023regularCS}; and
\textit{``Yeah, I am channeling Abby ... and I am not having fun with this program''}~\cite{burnett2016field}.

\boldification{**more NOT-TO-USE reasons: ultimately, intuition }

But perhaps the most important benefit of  manually <x>Mag'ing is to build and use intellectual ideas and devices that facilitate better design.
For example, experience using the <x>Mag manually has led to common vocabularies for discussing design and implementation choices~\cite{burnett2017microsoft}.
Another outcome has been <x>Mag ``Moments''---i.e., using an <x>Mag for just a moment, to consider how a particular <x>Mag persona would react if a feature behaved in a different way~\cite{hilderbrand2020engineering}. 
As is well-known in education, such intellectual devices require putting in the labor to build the expertise to use them~\cite{kolari2008learning}.
Stated another way, the cost of less effort is less expertise.

Ideally, then, practitioners would use both. 
As Section~\ref{subsubsec:hitsAndMisfires} showed, running the tool and a manual session on the same scenarios catches the most bugs. 
But when doing both is not viable, the trade-offs above suggest when to lean which way. 
We recommend:
\begin{enumerate}
    \item Lean toward Apply-<x>Mag when time is short, to replace environmentally expensive <x>Mag'ing over zoom, or for the broadest per-screen coverage---a fast, high-precision pass across whole screens. 
    \item Lean toward doing <x>Mags manually to analyze information-seeking sequences in workflow, since scent-following is where the tool was weakest.
    \item Lean toward manually <x>Mag'ing when finding bugs is not the only goal, since the sociotechnical benefits above come from doing the work socially and effortfully. 
\end{enumerate}

A team could sometimes run Apply-<x>Mag to reap the benefits of item~(1) above, and hold occasional manual sessions for the benefits of (2) and (3).
We recommended matching decisions as to when to use the tool vs. working manually based on the circumstances in (1)--(3), for a mix that draws upon the best of both worlds.



\subsubsection{What about the rest of InclusiveMag?\draftStatus{D2.7 (FAM 9/10 11:50pm)}}

Apply-<x>Mag automates only Step-3 of InclusiveMag (Figure~\ref{fig:inclusivemag}), the step practitioners do. That leaves Steps~1 and~2, the steps inclusive design researchers do to create an <x>Mag in the first place. Should those steps be tooled up as well?

InclusiveMag's Step-1 (Scope) requires two decisions:  technology type and diversity dimension. 
We caution against automating either one. 
Choosing a diversity dimension commits a researcher to a population and to extensive work on that population's behalf, and choosing a technology type commits them to a domain they must know well enough to argue about. 
These decisions involve commitments and skills that genuinely reflect a researcher's interests and expertise, not outputs of an LLM. 

The rest of Step-1 is research and analysis, and here AI could potentially help. 
We believe the researcher should do this work---read  literature on the dimension, investigate attributes differing substantially across it, etc.---and be able to defend these decisions later. 
But AI could provide pointers to additional sources the researcher had not found, could cluster attributes the researcher has gathered into candidate facet types, and could critique the researcher's tentative facet choices, such as flagging endpoints that sit too close together to span a useful range, or value ranges that are not approximately ordinal (Section~\ref{subsec:background}).

Step-2 (Derive) splits the same way. Constructing the personas from the facet values is regular enough that AI could produce a first cut without detriment to the researcher's intellectual engagement.
As Section~\ref{sec:background+Related} explained, the facet values completely define each persona, so turning them into a coherent persona is more composition than thinking, and the researcher can refine it as needed. 
Specializing the analytic process is different. 
The heuristics or walkthrough questions are where the researcher's vision translates to artifacts future practitioners must apply, and we would want a researcher to create them, with AI potentially critiquing what \added{they create.}

The pattern in both steps is that decisions should stay with the researcher, whereas critiquing and helping improve could be assisted by AI. 
This is the same trade-off as discussed above, one level up. There, practitioners who let the tool do \textit{all} their <x>Mag'ing save time but give up the expertise that manual sessions build. 
Here, researchers hypothetically letting an LLM scope and derive their <x>Mag would save time too, but the resulting <x>Mag would rest on expertise none of the <x>Mag creators actually had.

\subsection{Limitations\draftStatus{top D3 (it's empty!) MMB 9/8/26: 1306) }}
\label{sec:threats}

\subsubsection{Limitations of the evaluation\draftStatus{D2 (MMB 9/9)}}

As with any empirical study, our evaluation has limitations. 

\boldification{1a: generality}

The generality of our results may be limited to data we were able to obtain.
To mitigate this threat, we evaluated with multiple  technology products (7), contexts (2), countries (2), practitioners (6 teams + 2 instructors), and ultimate target audiences (myriad).
However, our field data lacked a few screens that humans reported to be bug-free.  
Also, it was not possible to evaluate Apply-<x>Mag's generality across every possible field situation, and future research is needed to further establish generality.

\boldification{** 1d: participant validation}
Human evaluators served as our ground truth, but  human evaluators' attitudes, biases, and social dynamics could have influenced the evaluation's results. 
We attempted to mitigate such social threats by counting inclusivity bugs using union (as versus consensus) of humans' bug-finding.  
That is, if anyone on a team said something was an inclusivity bug, we counted it (as in  other works~\cite{burnett2016field, busteed2026}).
We further mitigated this threat to our ground truth with participant validation by the products' producers.

\boldification{**1b: only 2 Mags + intersect. Also no UI}

We evaluated Apply-<x>Mag using only SESMag and GenderMag. 
We chose those two because using both versions manually have previously been field-tested and their manual application has demonstrated high precision \cite{burnett2016gendermag, busteed2026, guizani2022whywherefix}, providing strong comparison points for evaluating the tool. 
Still, our results do not establish that Apply-<x>Mag will achieve comparable effectiveness for every existing or future <x>Mag. 
Also, we evaluated only precision, recall, and environmental costs, but participants never sat down in front of Apply-<x>Mag, so we did not evaluate its user interface.
Evaluating additional <x>Mags, and evaluating participants' user experience with Apply-<x>Mag remains important future work.

\boldification{**1c: 1 persona.}

Also, recall from Use-case~1 that, for inclusivity, evaluating personas for both endpoints of each facet type is needed (e.g., for both risk-averse and risk-tolerant users).
Our investigation instead used only traditionally under-served facet values; i.e., SESMag's Dav (facet endpoint values statistically common among lower-SES users) and GenderMag's Abi (facet endpoint values statistically common among women).  
This does not measure the tool's effectiveness for target populations who are already reasonably well-served.
Evaluating the tool’s effectiveness across additional personas remains future work.

\boldification{**1f: environment }

Our environmental comparison carries two limitations.
First, the results are GreenQuery estimates, which require a browser session, so are not measurements of the actual tool because it needs an API to run  GPT-5.1.
This was necessary to compare the cost of the tool's detection-tactic and analysis prompts on an equal footing with a human chatting with GPT.
Second, we measured with one course's two-screen scenario, using zero-shot configurations (i.e., no worked examples).
We also put everything necessary into one human-chat prompt instead of assuming iterative clarifications, which minimized the human-chat cost.
Obviously, different chat scenarios,  different scenarios, different screens, and different sizes of <x>Mag configurations would produce different results.
That said, it was a conservative measurement, designed to make human chatting as environmentally competitive as possible against the tool's environmental costs. 

\boldification{**1g: model, params, generative variability}

Finally, Apply-<x>Mag uses an LLM, and LLMs have generative variability, so results may vary from those shown here.
Other factors that could change the results depend on which LLM are run with which parameters.
Our evaluation used GPT-5.1 with only its default API parameters to reduce the risk of evaluation-specific optimization. 
Using different models and/or parameters could produce different results.

\subsubsection{Limitations of the tool\draftStatus{MMB 9/8/26 13:00: D2.2}}

Apply-<x>Mag itself also has limitations---one inherent to the tool, some from its dependencies, and some with its current implementation.
The inherent limitation is that it performs its work via specialized heuristic evaluations (<x>Mag-HEs), which is different from the more common InclusiveMag practice of using specialized cognitive walkthroughs (<x>Mag-CWs).
<x>Mag-HEs' strength is breadth: <x>Mag-HEs look at all of each screen, whereas <x>Mag-CWs instead follow only a specific linear path through screens. 
The weakness is Apply-<x>Mag's limited capabilities with sequences of scent-following, as noted earlier in Section~\ref{subsubsec:hitsAndMisfires}.
That said, Apply-<x>Mag's precision and recall were higher than previous tools using <x>Mag-CWs, suggesting that the strength may outweigh the weakness.

In the dependency category, Apply-<x>Mag is only as good as the assets it harnesses.  
If the harnessed LLM is/becomes  ineffective, Apply-<x>Mag cannot be effective either.
Likewise, Apply-<x>Mag depends on the quality of an <x>Mag's configuration---solid foundations behind the facet types, associated heuristics, and if provided, the worked examples.  
If the <x>Mag behind the scenes is not solidly evidenced, Apply-<x>Mag'ing it cannot make it better.

Finally, our implementation of Apply-<x>Mag is currently alpha-level (proof of concept).
It does not yet globally store, because it has no safeguard against an <x>Mag's stored configuration being overwritten; future versions will globally store with safeguards, such as offering password protection of <x>Mags and/or pull request mechanisms. 
Also, it has a known bug in automatically ``union'ing'' its intersectional bug reports.
Still, we provide Apply-<x>Mag's code in the Supp. Docs. for interested readers to try out.


\section{Conclusion\draftStatus{d2.75}}
\label{sec:conclusion}

Designing technology inclusively is important: it can increase the size of a product's market, improve users' experiences, and do societal good by reducing the digital ``have's'' vs. ``have-nots'' chasm.  
Yet, inclusive design's labor intensiveness can make it prohibitively expensive for some HCI practitioners.

To address this problem, this paper presents Apply-<x>Mag, a tool to help HCI practitioners find and fix inclusivity bugs.  
Apply-<x>Mag is different from other HCI tools: 
\begin{itemize}
    \item \textit{General}: Apply-<x>Mag is general enough to support any <x>Mag---any inclusive design method that can be expressed in terms of facet value ranges and heuristics.
    \item \textit{Flexible}: Apply-<x>Mag is flexible enough to use full methods, subsets, and combinations of these, enabling intersectional insights (\Cref{subsec:Use-case4,subsec:results-Use-case4}).
    \item \textit{Practical viability}: It does not expect practitioners to enter information they might not have (\Cref{subsec:Use-case2,subsec:results-Use-case2}). Further, it avoids persona explosion, or even designing any personas at all, because its personas are just a collection of facet endpoint values.  Given this, evaluating with just 2 personas finds inclusivity bugs for a wide range of users (\Cref{subsec:Use-case1}).
    \item \textit{Fidelity/Effectiveness}: Apply-<x>Mag confirmed 97 of the bugs humans found using the \textit{same} <x>Mags manually, plus 32 more bugs that humans missed but later confirmed, and missed only 18. Overall, its precision ran 90--99\% and recall 82--89\% (Section~\ref{subsec:results-Use-case1}).
    \item \textit{Environmentally responsible}: Each screen evaluation cost about the same resources as 2--4 ordinary (non-AI) Google searches (Section~\ref{subsec:results-environment}).
\end{itemize}

Perhaps its most important attribute is that Apply-<x>Mag builds on the shoulders of research giants.
Its generality facilitates practitioners applying \textit{any} inclusive design methods that previous researchers have created---and even methods created in the future.
By putting this growing body of research within practitioners' reach, Apply-<x>Mag makes inclusive design itself more accessible.

\added{
\textbf{Data availability}: The supplemental materials are available in \url{https://doi.org/10.5281/zenodo.22678394}.
}

\begin{acks}
We thank all the participants for their time and insights. We also thank Alec Basteed and Katie Kimura for their contributions to preparing this paper, and MOSIP for helping with our studies.  This work was supported in part by NSF grants.

\end{acks}

\bibliographystyle{ACM-Reference-Format}
\bibliography{bib}

\end{document}